\documentclass[letterpaper,amsmath,showpacs,amssymb,nofootinbib,10pt]{revtex4}
\usepackage{color}
\usepackage[dvipsnames]{xcolor}
\usepackage{newlfont}
\usepackage{graphicx}
\usepackage{amssymb}
\usepackage{amsmath}
\usepackage[caption=false]{subfig}
\usepackage{booktabs}
\usepackage{bm}
\usepackage{verbatim}
\usepackage[latin1]{inputenc}
\usepackage{bbm}
\usepackage{multirow}
\usepackage[thinlines]{easytable}
\usepackage{braket}
\usepackage{amsthm}
\usepackage{bbold}
\usepackage{subfig}
\usepackage{mathrsfs}
\usepackage[varg]{txfonts} 
\usepackage{amsfonts}
\usepackage{comment}

\usepackage{geometry}
\usepackage[pagebackref=false, colorlinks=true]{hyperref}
\definecolor{redish}{rgb}{0.7,0.2,0.0}  % color defined in (r=red,g=green,b=blue) model
\definecolor{bluish}{rgb}{0.2,0.5,0.8}
\hypersetup{linkcolor=red,          % color of internal links
	citecolor=magenta,        % color of links to bibliography
	filecolor=magenta,      % color of file links
	urlcolor=teal}          % color of the url
\newcommand{\ba}{\begin{array}}
	\newcommand{\ea}{\end{array}}

\begin{document}
\title{Loss-aware state space geometry for quantum variational algorithms}
\author{Ankit Gill}\email{ankitgill20@iitk.ac.in}
 \affiliation{Department of Physics, Indian Institute of Technology, Kanpur 208016, India}	
\author{Kunal Pal}  \email{kunal.pal@apctp.org} 
\affiliation{Asia Pacific Center for Theoretical Physics, Pohang 37673, Republic of Korea}

%\date{February 2026}
\begin{abstract}
    The natural gradient descent optimisation technique is an efficient optimising protocol for broad classes of classical and quantum systems that takes the underlying geometry of the parameter manifold into account by means of using either the Fisher information metric of the classical probability distribution function or the Fubini-Study tensor of the associated parametrised quantum states in the consequent update rules. Even though the natural gradient descent procedure utilises the geometry of the space of probability or states, it is, however, insensitive to the measure of parametrised distance on the space of possible outcomes when the corresponding  optimising problem is considered for the expectation value of a classical or quantum observable with respect to the probability distribution or the quantum state. In this work, we introduce a generic optimising principle, where the intrinsic geometry of the space of outcomes has been taken into account suitably, either by using an ambient space construction with a base statistical manifold with the usual Fisher information metric (or the Fubini-Study tensor), where the loss hypersurface is embedded to, or by means of a first-principle construction from the overlap of nearby quantum states on the projective Hilbert space. This construction as well as a family of conformal variants yields a form of loss-aware natural gradient updates that rescale the effective step size while preserving the descent direction. We benchmark the resulting optimisers on  variational quantum circuit examples and on a classical neural network task, finding that, while the standard natural gradient remains the most robust on average, the proposed conformal schemes can improve best-case convergence in favourable regimes. We subsequently develop a biorthogonal formalism that leads to gauge-invariant kernel-weighted geometry, resulting in a formalism governed by a loss-adaptive non-Hermitian tensor structure, as well as a pair of non-metric connections, analogous to the $\pm\alpha$-connection of classical information geometry.
\end{abstract}

\maketitle

{\hypersetup{linkcolor=gray}
\tableofcontents
}
\section{Introduction}

The quantum variational eigensolver (QVE) is a powerful classical-quantum hybrid algorithm that is suitable for current  noisy
intermediate-scale quantum (NISQ) computers \cite{Eisert2025mind}. The QVE, formulated in \cite{Peruzzo2014variational}, generally relies on preparing a parametrised quantum state and a cost associated with such a state, which in most cases is the measured value of a (Hermitian) operator \cite{Tilly2022variational}. Then a classical optimiser is used to update the circuit parameters so that the value of the cost is lowered for the upcoming iteration. This idea broadly \cite{Mcclean2016theory}, and various generalisation of the same, as well as formulation of similar motivations have been used widely in the last decade or so in various contexts ranging from applications in the nonlinear Schrodinger equations \cite{Lubasch}, to analyse the spectrum of mass operators in black holes and cosmological background in \cite{Joseph2022quantum}, to noisy variational quantum circuits in \cite{Fontana:2023wvl, Shao:2023xki}, to  probe nuclear structure \cite{Romero:2022blx, Carrasco-Codina:2025gan}, in lattice gauge theories \cite{Popov:2023xft}, for $\text{SU}(N)$ Fermions in \cite{Consiglio:2021upq} for multiparameter estimations in \cite{Cimini:2023uri}, for quantum systems in finite temperature in \cite{WuPRL, Selisko:2022wlc}, for relativistic systems in \cite{Chawla}, for quantum many-body systems in \cite{Araz:2024xkw, Medina:2023axo, Alvertis:2024lxc}. These mentioned works, of course, refer only to a very selective few of the large number of works that appeared over the years \cite{Gidi2023stochastic, Smart2024many, Miura2026velocity, Rogerson2024quantum, Malvetti2024randomized, Vogl2025variational, Wiedmann2025convergence, Li:2025kod, Joch:2025rsc, Patra2025resource, Choudhury2025fragment, Clemente, Barligea, Sherbert, Wang:2025xlq, Okada, Jattana, Sewell, Kim2024qudit, Lee2026distributed, Satoyuki, Nakayama} and a proper review is out of the scope of the current work; for this, we refer the reader to the reviews \cite{Tilly2022variational, daSilvaFonseca:2025jtk, Callison:2022zfe, Symons:2023fgo, Kyaw:2022pip, Cerezo, Yao, Watanabe}, where detailed discussion on the various aspects and citations of the related literature can be found.

One of the key aspects for the efficient performance of this algorithm is the awareness of the intrinsic geometry of the parameter manifold, which is typically not included in ordinary gradient descent (OGD)-type algorithms. If the parameter space in question has a non-trivial geometry, induced by the corresponding probability distribution function (PDF), then it is expected that different directions of the parameter manifold will have anisotropic cost associated with traversing that particular direction, which in turn will have a non-trivial impact on the optimisation procedure in that manifold. To this end, one of the most powerful geometry-aware gradient optimisations has been argued to be the one that uses the natural metric on the parameter space, namely the unique Fisher information metric (FIM) of such a space of classical PDFs \cite{Amari1998, Quinn2023information}. For a PDF, parametrised by a set of parameters, the parameter manifold can be endowed with a unique metric structure (as well as a pair of mutually dual connections), which essentially controls the direction of the so-called \textit{natural} gradient descent (NGD) on such a manifold \cite{Amari2000methods}. The NGD typically shows better convergence properties to the local minima than that of the OGD, and has been used in various contexts \cite{Amari1996neural, Rattray1998natural,  Amari2019fisher, Liu2025efficient, Patel2025natural, Miyahara2025information}.

For quantum systems similarly, the canonical inner product on the Hilbert space naturally induces a notion of geometry on the complex projective space $\mathbb{C}P^n$ of the pure quantum states, which is described by the (Hermitian) Fubini-Study (FS) tensor on such a complex manifold. For parametrised quantum states, parametrised by a set of real parameters, the geometry then is governed by the pull-back of the FS tensor on the parameter (sub-) manifold, which consists of the real as well as the symmetric metric (QMT) and the purely imaginary and anti-symmetric $2$-form, known as the Berry curvature in the literature, which are by construction invariant under the $U(1)$ rescaling of the quantum states which is essential for it to be physically meaningful. The natural gradient descent algorithm that uses this geometry of the quantum states, dubbed as the quantum natural gradient (QNG) was formulated in \cite{Stokes}, which structures the variational problem with a block-diagonal approximation of the QGT of the associated circuit. Importantly, it was shown that this QNG is more efficient than that of the standard gradient descent in terms of the rate of the convergence of the algorithm. The performance of the QNG in comparison with the OGD, as well as the connection of the same with the imaginary-time evolution generated by the Hamiltonian have seen a flurry of research activities \cite{Yamamoto, Kolotouros2024random, Shi2026weighted, Dell2025quantum, Roy2024efficient, Minervini2025quantum, Koczor2022, Haug2022natural}.

Even though the FIM with the associated NG, as well as the QNG, exploits the geometry of the (in general curved) space of parametrised PDFs, it is not sensitive to the `geometry' of the cost function itself, which for quantum circuit optimisation is typically taken to be the expectation value of a Hermitian operator (with respect to the canonical inner product on the Hilbert space) in the parametrised states. To elaborate, the cost function, which we will denote from now on as $L$, defines a co-dimensional-one hypersurface in the ambient space, which in this case is the product manifold $\widetilde{\mathcal{M}} = \mathcal{M} \times \mathbb{R}$ and the metric on the ambient space induce a geometric structure on this hypersurface. In the context of the optimisation algorithms, the Riemannian metric induced on the hypersurface by this embedding map can then be used to design (possibly efficient) series of optimising algorithms. One such  optimiser was used very recently in \cite{Harvey2025optimiser}, for neural network training, where the induced metric was used as a preconditioning method and the efficiency was found to be improved than that of the available and popular methods; particularly in the low-dimensional models studied in \cite{Harvey2025optimiser}.  These results provide not only a set of new, practical and cost-efficient optimiser methods, but also pave the way for further rigorous  studies on how to incorporate the geometry of the objective-space into building optimiser algorithms. Another conceptually pleasing aspect of the optimiser used in \cite{Harvey2025optimiser} was that the modified NG techniques have a natural interpretation in terms of a form of gradient clipping technique, namely, the (effective) learning rate is decreased in the  regions with high curvature, thereby providing a `natural' cut-off for the corresponding gradient stability; on the other hand, it does keep the larger learning intact in the lower curvature regimes.  The geometry induced by the pull-back metric thus provides a way to control the effective step-size in a loss-aware manner and, in principle, can be tuned to design further possibly-efficient optimisers. On the other hand, it would be desirable for any such algorithm to have a  computational complexity similar to that of traditional variational ones. Satisfactory formulations of potential loss-aware optimisers thus require to strike a subtle balance between these two points; a perspective we want to investigate in the present work, specifically from an information geometric point of view \cite{Amari1998}. 

We propose and study a class of loss-aware pulled back geometry motivated by the work of \cite{Harvey2025optimiser}, where we embed the cost-dependent hypersurface in the space of parametrised PDF, with the geometry of the base manifold being governed by the FIM of the associated PDF, both classical and quantum  (with proper modification), aiming to provide any possible analytical and  computational advantage without sacrificing the stability of the same. To this end, we first explore the ambient space construction based on the embedding of a cost function(al) in the space of (classical) PDFs with a suitable parametrisation, where the geometry is governed by the unique FIM and a pair of dual $\pm\alpha$ connections \cite{Amari2000methods}. We construct the induced metric and the NGD problem is governed by this pull-back metric, which is essentially a rank-$1$ deformation of the FIM, and does not change the direction of the steepest descent; only modifies the effective step-size of the update rule. 

For a parametrised quantum system, on the other hand, we embed the cost-hypersurface in a space where the geometry is governed by the QMT of the associated (pure) state-space and the embedding induces a metric on the hypersurface which is gauge-invariant by construction and hence is physically meaningful. We develop a gradient optimiser scheme based on this induced metric, which we refer to as a loss-aware (quantum) natural gradient (LA-QNG) optimiser, which is similar to \cite{Harvey2025optimiser}, can be thought to be equivalent of the gradient clipping technique, which is in-built with the geometric construction itself. We perform the LA-QNG optimising task with a quantum circuit and the corresponding block-diagonal approximation of the QMT \cite{Stokes} and compare the performance of LA-QNG with both OGD as well as QNG. 

In order to have more control over the effective learning rates, we next propose a conformal modification of the induced metric, which zooms in (or out) the geometry while keeping the intrinsic angles invariant and, in our opinion, complements the pure LA modifications, which we note that scales different directions anisotropically. The conformal transformation of the induced geometry acts as an overall scale transformation of the geometry and hence changes the effective step-size, without affecting the direction of the NG. 

To assess if these geometric modifications of the  NG approaches can deliver any practical advantages in the optimisation problem over the existing ones, we have tested our proposed method in variational quantum circuits with the QGT constructed using the block-diagonal approximation. Our numerical analysis suggests that, even though the QNG provides the best convergence overall, one of the conformal variants (CLA-$3$-QNG) can deliver superior performance in favourable circumstances. We have also performed a classical optimiser test with the base metric being the FIM, determining the preconditioning updates rules. In this setting, the (classical) conformal variant (CLA-3-NG) again achieves the best performance, surpassing other standard FIM-based update rules such as the natural gradient, as well as methods like Adam or SGD-RMS \cite{Harvey2025optimiser}.

Even though in the preceding section we have outlined how to obtain the LA-FIM or LA-QMT from the pull-back of the ambient space metric on the loss hypersurface, it is not a priori clear if such type of metrics  can be induced on the statistical manifold under consideration from a physically meaningful divergence function. At this point, we remind the reader that the traditional construction of the classical information geometry is based primarily on a well-defined divergence function, which induces a Riemannian metric in the second-order expansion and a pair of dual connections through the third-order properties \cite{Eguchi1992geometry}. The well-known divergence functions, such as the Kullback-Leibler (KL) divergence \cite{Kullback1951information} and the $\alpha$-divergence \cite{Amari1982differential, Zhu1995bayesian}, which  typically does not represent distance between PDFs, have been used widely in the literature to provide the geometric structure on the statistical manifold in a rigours way. This motivates us, in particular, for parametrised quantum states, to investigate in what extent a LA-geometric structure can be formulated from a first-principle analysis of overlap-distance between `nearby' pure quantum states, as was done for example in the standard QGT construction in \cite{Provost1980riemannian}. To this end we use the polar decomposition of the (position-space) wavefunction (for continuous variable quantum systems) to formulate a geometric structure, which helps us to build an intuitive exploration in parallel with the classical formulation.  We need to emphasise two points here about the construction that we will describe in the sequel; first of all, even though the use of the position-space wave function provides a form of the QMT (as well as the associated Berry curvature) that is at a similar status as that of the FIM \cite{Facchi2010classical}, with the fixed normalisation of the wavefunction, it is not possible to obtain a non-trivial (non-metric) connection on the statistical manifold, unless we use two mutually biorthogonal functions. Secondly, the LA-geometry with the cost as the expectation value of a Hermitian operator for the constructed parametrised circuit will impose a non-local structure in the geometry, both in the QMT as well as in the analogue of the two $\pm\alpha$ connections. We describe in detail the biorthogonal construction of the Hermitian structure from such an overlap, which can be decomposed into four independent contributions, where apart from the standard QMT and Berry curvature, we will also have two `flipped' contributions, namely, symmetric \textit{but} purely imaginary as well as anti-symmetric \textit{but} real tensors. We will also provide details of $\pm\alpha$-connections, which by construction are non-metric and thus give a complete description of the associated LA-geometry.

The paper is structured as follows: in section \ref{NG}, we briefly recapitulate the basics of information geometry in classical and quantum settings, which also sets up the notation we intend to use in the rest of the work. In the next section \ref{LA-QMT-1}, we explore the ambient space construction and the formulation of the subsequent induced metric in the variational landscape. Then we elaborate on our construction of the conformal class of LA-geometries in section \ref{sec:CLAgeometry}, which includes three different possible cases. We compare the performance of different types of geometries numerically in section \ref{secperformance} and discuss the possible advantages of using the CLA-type geometries. Exploration of the LA-geometries from the point of view of information geometry is done in section \ref{secdivergence}, where we propose a novel form of the LA-geometries induced on the statistical manifold from the systematic expansion of the overlap function of the states with nearby parameter values. To this end, we establish the position-space form of these geometries, which points towards essential non-local features of the associated QMT and the Berry curvatures for a generic operator kernel. The biorthogonal construction of the overlap-integral and the subsequent formulation of LA-$\alpha$-QMT and LA-$\alpha$-connections. In particular, we emphasise why now the geometric structures are non-Hermitian, and they will have novel flipped contributions to the geometry apart from the standard (Hermitian case) ones. Finally, in the appendix \ref{NGcomparisonexponential}, we compare the geometry associated with all the five classes of metrics discussed in this work in terms of the coordinate invariants of such geometries as well as the NG-trajectories of different metrics considered in this work for a simple toy model belonging to the exponential family of PDFs, whenever analytical results are possible to obtain.

%%%%%%%%%%%%%%%%%%%
\section{Natural gradient descent optimiser}\label{NG}
In this section, we will, in brief, describe the basic ingredients of information geometry for the classical parametrised PDFs and also explain how this is related with the natural geometry on the projective space of the pure quantum states, governed by the FS tensor. We will consider a general family of PDF, which we will denote as \( P(x; \theta) \) for a random variable \( x \),  parametrised by a set of continuous parameters $\{ \theta\}$, where the notion of distinguishability between two such distributions is quantified by the classical Fisher information metric (FIM)
\begin{equation}
g^{\text{FIM}}_{ij}=\mathcal{E}_{p}\Big[\partial_{i} \ln  P(x;\theta)~\partial_{j}\ln  P(x;\theta)\Big]~.
\label{Fishermetric}
\end{equation}
Here and in subsequent discussions, we use $\mathcal{E}_{p}[\cdot]$ to represent the statistical average of a quantity with respect to the PDF under consideration, and the partial derivatives are with respect to the parameters $\theta^{i}$ \footnote{We will always denote the coordinate index with a superscript like $\theta^{i}$, which follows different transformation laws in a curved manifold than that of $\theta_{i}$.}. The FIM represents a unique metric in the space of parametrised PDFs, which can be obtained as a second-order contribution to the geometry from the expansion of the standard Kullback-Leibler divergence; which also provides the dual $\pm\alpha$-connections  as third-order contributions \cite{Amari2000methods}. The natural gradient descent  technique (NG) uses this geometrical structure on the relevant statistical manifold, where the update direction is governed by the Riemannian metric in the manifold, and can perform significantly better than the standard gradient descent ones, which uses the Euclidean norm to update the optimiser \cite{Amari1998}. The NG thus updates the gradient direction as 
\begin{equation}
    \theta_{t+1}^{i}=\theta^{i}_{t}-\eta g^{(\text{FIM})ij}\partial_{j}L(\theta)~,
\end{equation}
for a chosen step-size $\eta$, which shows that for optimisation, the point $\theta^{i}_{t}$ must move in the \textit{opposite} direction to the natural gradient of the loss function with respect to $ g_{ij}^{\text{FIM}}$. Even though the NGD most often performs better than that of the SGD, it should be noted that the FIM inherently does not contain any information about the sample space; it solely represents the geometry of the parameter manifold for a chosen parametrisation of the PDF.

For quantum systems described by a family of states depending on a set of parameters, a notion of distance between states can be introduced in the space of quantum states or, more generally, in the space of density matrices \cite{Provost1980riemannian}, which for pure states naturally arises on the complex projective Hilbert space. In this space, one can define a Hermitian tensor that is induced by the Hilbert space inner product, known as the Fubini--Study (FS) tensor. For a quantum state $\Psi$, assumed to be normalised to unity, the FS tensor is invariant under global $U(1)$ phase transformations, which makes it a physically-meaningful measure of distance \cite{Brodygeometric}. When the FS tensor defined on the complex projective space is pulled back to the manifold of parameters characterising the quantum state, one obtains the quantum geometric tensor (QGT). Expressed in terms of real coordinates $\{\theta^{i}\}$ that parametrise the pure state, the tensor takes the form \cite{Ashtekar, Kibble1979, Braunstein94, Field1999geometry, Anandan1991}
\begin{equation}
\text{FS}_{ij}=
\braket{\partial_{i}\Psi(\theta) | \partial_{j}\Psi(\theta)}
-
\braket{\partial_{i}\Psi(\theta) | \Psi(\theta)}
\braket{\Psi(\theta) | \partial_{j}\Psi(\theta)}~.
\label{tensorstructure}
\end{equation}
Starting from this Hermitian tensor on the projective Hilbert space of pure states, it was shown in \cite{Facchi2010classical} that the real, symmetric component of the FS tensor can be written explicitly in coordinate form as
\begin{equation}
g^{\text{FS}}_{ij}
=
\frac{1}{4}\mathcal{E}_{p}
\Big[
\partial_{i}\ln P(x;\theta)\,
\partial_{j}\ln P(x;\theta)
\Big]
+
\mathcal{E}_{p}
\Big[
\partial_{i}\Phi(x;\theta)\,
\partial_{j}\Phi(x;\theta)
\Big]
-
\mathcal{E}_{p}
\Big[
\partial_{i}\Phi(x;\theta)
\Big]
\mathcal{E}_{p}
\Big[
\partial_{j}\Phi(x;\theta)
\Big] ,
\label{FSmetric}
\end{equation}
which is commonly referred to as the quantum metric tensor (QMT). The imaginary antisymmetric part of the FS tensor, on the other hand, defines a closed $2$-form on the projective Hilbert space,
\begin{equation}
\omega_{ij}
=
\frac{i}{2}
\mathcal{E}_{p}
\Big[
\partial_{i}\ln P(x;\theta)\,\partial_{j}\Phi(x;\theta)
-
\partial_{j}\ln P(x;\theta)\,\partial_{i}\Phi(x;\theta)
\Big] ,
\label{berrycuravture}
\end{equation}
which is known in the physics literature as the Berry curvature. These two-forms together provide a symplectic structure on the manifold. In writing the above expressions we have used the polar decomposition of the wave function in the position representation,
\begin{equation}
\Psi(x;\theta)
=
\sqrt{P(x;\theta)}\, e^{i\Phi(x;\theta)} ,
\end{equation}
where $P(x;\theta)$ and $\Phi(x;\theta)$ are two real functions for a state that depends on a set of $n$ parameters $\theta^{i}=(\theta^{1},\theta^{2},\dots,\theta^{n})$. It is assumed that the wave function, and therefore the functions $P(x;\theta)$ and $\Phi(x;\theta)$, are smooth and differentiable across the entire parameter manifold \footnote{The non-analyticity of the geometric quantities of the parameter manifold are of specific importance for quantum systems exhibiting ground-state quantum phase transitions, see, for example, the discussions in \cite{Zanardi2006, zanardi2007information, Dey, TS2, Jaiswal:2021tnt,  Streleck}.}.
It is important to note that the metric tensor in Eq.~\eqref{FSmetric}, introduced in \cite{Facchi2010classical}, differs from the metric structure obtained for a classical probability distribution even when both share the same probability density $P(x;\theta)$. This difference arises due to the presence of the non-trivial phase $\Phi(x;\theta)$ of the quantum wave function, which encodes genuine quantum effects. The Berry curvature can alternatively be interpreted as the field-strength tensor associated with the Berry connection
$A_i = i \braket{\Psi(\theta) | \partial_i \Psi(\theta)} $,
such that
$F_{ij} = \partial_i A_j - \partial_j A_i $,
which is manifestly antisymmetric in its indices \cite{Berry1984}. 

In a similar motivation to classical NGD, the optimisation problem for a cost function, when the geometry is governed by the QGT was formulated in \cite{Stokes}, where the superiority of the QNG against the traditional  optimisers was established, using a block-diagonal approximation of the QGT. Again we note that, similar to the classical counterpart, the QGT also is explicitly insensitive to the form of the cost function being used, which for quantum systems, can be traced back to the nature of the operator-kernel being used to define the cost at the first place. The fact that any refined information about the geometric structure of the cost function should be advantageous to optimise the same primarily motivates us to introduce the loss-aware geometry in the classical and quantum systems in the subsequent sections.

%%%%%%%%%%%%%%%%%%%%
\section{Loss-aware pull-back Geometry in classical and quantum systems} \label{LA-QMT-1}
In this section,  we will present a new optimiser algorithm that takes into account the geometry of the loss landscape in a possibly efficient way, following the recent work of \cite{Harvey2025optimiser}. In this paper, the author considered the embedding of the loss landscape in a higher-dimensional ambient space and developed the optimiser based on the induced metric on that hypersurface, which shows better performance than other optimisers, specifically for low-dimensional models considered.
Motivated by this work, here, first we will consider a natural gradient algorithm, where the governing geometry is induced by the pull-back of the standard Fisher information metric (FIM)  of the probability distribution, which in turn is embedded in a higher-dimensional ambient manifold, where the embedding function is controlled by the loss function of the problem. Similarly, for quantum systems, we will consider the embedding of the loss function in a higher-dimensional product manifold of the standard parameter manifold (with the QMT determining the distance) augmented with a Euclidean direction. Finally, in the subsequent subsections we will provide the position-space form of the induced metric on the hypersurface, and how to use it as a NGD algorithm.

\subsection{Ambient Space Construction and the pull-back metric on the classical parameter manifold}

Let us consider a generic statistical manifold $\mathcal{M}$ with the usual geometric structures, most importantly the FIM, $g^{\text{FIM}}_{ij} (\theta)$, on it, where we have denoted the collective coordinates as $\{\theta\}$. Next, we define an extended manifold of the form $\widetilde{\mathcal{M}} = \mathcal{M} \times \mathbb{R}$, with coordinates
$X^A = (\theta^1,\ldots,\theta^n,z).$
Here we introduce an ambient metric of block-diagonal form
\begin{equation}
	\widetilde g_{AB}(\theta)
	=
	\begin{pmatrix}
		g^{\text{FIM}}_{ij}(\theta) & 0 \\
		0 & 1
	\end{pmatrix},
\end{equation}
which implies that the ambient line element is of the form: 
\begin{equation}
	\text{d}s^2_{\text{ambient}}
	=g^{\text{FIM}}_{ij} (\theta)\, \text{d}\theta^i \text{d}\theta^j + \text{d}z^2.
\end{equation}

Then to capture the geometry of the loss-landscape, we define a hypersurface embedding of the variational manifold in the ambient space, where the embedding function is of the form 
\begin{equation}
	\phi : \mathcal{M} \hookrightarrow \widetilde{\mathcal{M}},
	\qquad
	\phi(\theta) = (\theta^1,\ldots,\theta^n, z = f[L(\theta)]),
\end{equation}
with an arbitrary functional $f[L(\theta)]$ of the loss function $L(\theta)$, which for simplicity we can assume to be a linear functional of the form $f[L(\theta)]=cL(\theta)$, for some real parameter independent constant $c$, which can be chosen to be unity by redefining the normalisation. For this embedding hypersurface, 
the induced metric on $\mathcal{M}$ is the pull-back
\begin{equation}
	g = \phi^{*} \widetilde g~,
\end{equation}
which in explicit coordinate notation have the form
\begin{equation}
	g^{\text{LA}}_{ij}=
	\widetilde g_{AB}
	\frac{\partial X^A}{\partial \theta^i}
	\frac{\partial X^B}{\partial \theta^j}.
\end{equation}
Using the particular form of the embedding functions in the ambient space, we obtain the explicit expression of the induced metric on the variational manifold as \cite{Harvey2025optimiser}
\begin{equation}
	g^{\text{LA}}_{ij}(\theta)=g^{\text{FIM}}_{ij}(\theta)+\partial_i L(\theta) \, \partial_j L(\theta),
\label{LA-metric}
\end{equation}
which now can be thought to be governing the geometry of the loss hypersurface. If the embedding is smooth, which is essential to find an inverse of the metric $g^{(\text{LA})ij}$, we will use the well-known  Sherman-Morrison inversion technique. The effective learning rate for the LA-NG, based on the induced metric and how it encompasses the gradient clipping naturally, is explained in \cite{Harvey2025optimiser} in detail, and these are also valid for the metric \eqref{LA-metric} also with proper modifications. However, we stress that the effect of embedding is essentially to `stretch' the metric $g^{\text{FIM}}_{ij}$ anisotropically depending on the tangent directions of the loss $L(\theta)$; it can even induce non-trivial non-diagonal components, even if the FIM is diagonal for a given choice of parametrisation. To elaborate, if we view the two variational manifolds equipped with $g^{\text{FIM}}_{ij}$ and $g^{\text{LA}}_{ij}$ as an `exact' change of the underlying geometry (apart from originating due to a coordinate transinformation), then, generally, the distance between two `same' points with respect to the metric $g^{\text{LA}}_{ij}$ is always greater than that with respect to $g^{\text{FIM}}_{ij}$. However, for the curves representing the image of the level curves on the parameter manifold, the distance measures along two points on such a curve, as measured by both  $g^{\text{FIM}}_{ij}$ and $g^{\text{LA}}_{ij}$ are the same.

%%%%%%%%%%%%
\subsection{Pull-back metric on the quantum parameter manifold }\label{classicalpullback}
In the quantum analogue of the LA metric, where the role of the FIM is played by the (real part of the) pull-back of the Fubini-Study (FS) tensor on the parameter manifold (known as the quantum geometric tensor), since the Hermitian FS tensor governs the natural symplectic structure on the projective Hilbert space of the pure quantum states, and is invariant under a $U(1)$ transformation, have the form of \eqref{FSmetric}.
Then we can follow the similar ambient-space construction of the classical case to consider  the induced metric on the hypersurface governed by the loss function $L(\theta)$ and can obtain a formally similar expression like the classical case \eqref{LA-metric} and we denote it by $g^{\text{LA}}_{ij}(\theta)=g^{\text{FS}}_{ij}(\theta)+\partial_i L(\theta) \, \partial_j L(\theta)$. Note that it is by construction invariant under a phase transformation of the parametrised state, a crucial property that such metrics must satisfy for it to be a physically relevant one. For convenience, we will refer to this kind of geometries of the parameter manifold of our interest as the loss-aware quantum metric tensor (LA-QMT) or LA metric from now on.
\paragraph{Position space representation:}
To understand the role of loss-aware correction in standard QMT, we perform a position-space representation of the quantum states and use the polar decomposition: $\Psi(x;\theta)=\sqrt{P(x;\theta)}e^{i\Phi(x;\theta)}$ and substitute it back into \eqref{LA-metric}. The contribution of the standard QMT is noted in the literature and is proportional to the quantum Fisher information of the associated state when the contribution from the pure-phase part vanishes \cite{Facchi2010classical}. Similarly, for the loss-dependent part, we obtain, using  full sets of complete position basis, 
\begin{equation}
\begin{split}
    \partial_i L(\theta) \, \partial_j L(\theta)=\int dx_{1}dx_{2}dx_{3}dx_{4}\sqrt{P_{1}P_{2}P_{3}P_{4}}e^{i\Phi_{21}}e^{i\Phi_{43}}\Big(\tilde{A}_{12}(\frac{1}{2}\partial_{i}\ln{P_{1}}-i\partial_{i}\Phi_{1})+\partial_{i}\tilde{A}_{12}+\tilde{A}_{12}(\frac{1}{2}\partial_{i}\ln{P_{2}}+i\partial_{i}\Phi_{2})\Big)\\
    \Big(\tilde{A}_{34}(\frac{1}{2}\partial_{j}\ln{P_{3}}-i\partial_{j}\Phi_{3})+\partial_{j}\tilde{A}_{34}+\tilde{A}_{34}(\frac{1}{2}\partial_{j}\ln{P_{4}}+i\partial_{j}\Phi_{4})\Big),
\label{eqlossposition}
\end{split}
\end{equation}
where we have used the notation $\tilde{A}_{ab}=\tilde{A}(x_{a}, x_{b})$, as well as $\Phi_{ba}=\Phi(x_{b})-\Phi(x_{a})$, and $P_{a}=P(x_{a})$, $\Phi_{a}=\Phi(x_{a})$.  This expression of the contribution from the loss to the induced metric \eqref{eqlossposition}, shows how non-locality inherently enters into the contribution of the effective metric and hence into the learning rate of the optimiser. Note that when the operator in question is the identity operator on the Hilbert space and the corresponding operator kernel is the delta function kernel, this term vanishes identically. In the generic of a Hermitian operator and a complex-valued (position-space) wavefunction, we can decompose the contribution from the cost term  \eqref{eqlossposition} as $\int dx_1\,dx_2\,dx_3\,dx_4\,
  \sqrt{P_1 P_2 P_3 P_4}\,
  e^{i\Phi_{21}}\,e^{i\Phi_{43}}\,
  \mathcal{B}^i_{12}\,\mathcal{B}^j_{34}$, where
$\mathcal{B}^i_{12} = \tilde{A}_{12}\!\left(\tfrac{1}{2}\partial_i\ln P_1 - i\partial_i\Phi_1\right)
+ \partial_i\tilde{A}_{12}
+ \tilde{A}_{12}\!\left(\tfrac{1}{2}\partial_i\ln P_2 + i\partial_i\Phi_2\right)$, which has contributions from a \textit{local} part
involving only the diagonal kernel $\tilde{A}(x,x)$ and a \textit{non-local} part
supported on $x_1 \neq x_2$. The local sector recovers a weighted classical Fisher information matrix and carries no
phase information, since the relative phase $\Phi_{21}$ vanishes on the diagonal; by
contrast, the non-local sector contains three physically distinct contributions: long-range
amplitude correlations $\partial_i\ln {P}_{1}\cdot\partial_j\ln {P}_{3}$ with $x_{1}\neq x_{2}$, a
purely quantum phase sector $\partial_{i}\Phi_{1}\cdot\partial_{j}\Phi_{2}$ that realises a
non-local generalisation of the Fubini--Study metric and is gauge invariant in the phase
difference $\Phi_{21} = \Phi_{2} - \Phi_{1}$, and cross amplitude-phase interference terms
encoding the quantum Fisher information. 

\paragraph{Pull-back metric for the exponential family:} Let us now consider one explicit example to see how the LA-contribution might change the  geometry induced by QMT. To simplify our analysis, we will assume the phase of the complex-valued wavefunction in the position space vanishes; then the QMT essentially boils down to the FIM of the associated PDF $P(x;\theta)$. On the other hand, the form of the LA-contribution to the pull-back geometry is 
\begin{equation}
  \mathcal{L}_{ij}=\partial_i L(\theta) \, \partial_j L(\theta)_{|_{0}}=\frac{1}{4}\int dx_{1}dx_{2}dx_{3}dx_{4}\sqrt{P_{1}P_{2}P_{3}P_{4}}\tilde{A}_{12}\tilde{A}_{34}\Big(\partial_{i}\ln{P_{1}}+\partial_{i}\ln{P_{2}}\Big)\Big(\partial_{j}\ln{P_{3}}+\partial_{j}\ln{P_{4}}\Big). 
\label{eq:classicalpullback}
\end{equation}
Even though this is a rather idealised situation, we will see that several non-trivial effects of the pull-back metric construction are already manifest in these analytical results. We will also assume that the PDF $P(x;\theta)$ associated with the quantum state belongs to a particular example of the exponential family; the normal distribution. This PDF can be written as a member of the exponential family, the generic form of the same in terms of  the canonical parameters $(\theta^{1}, \theta^{2})$ is \cite{Amari2000methods}
\begin{equation}
P(x;\theta)=\exp{\Bigg({C(x)+\sum^{n}_{j=1}\theta^{j}F_{j}(x)-\psi(\theta)}\Bigg)},
\label{ExponentialPDF}
\end{equation}
with $C(x)=0$, $F_{1}(x)=x$, $F_{2}(x)=x^2$, $\theta^{1}=\frac{\mu}{\delta^2}$, $\theta^{2}=-\frac{1}{2\delta^2}$ and $\psi(\theta^{1}, \theta^{2})=-\frac{(\theta^{1})^2}{4\theta^{2}}+\frac{1}{2}\log{(-\frac{\pi}{\theta^{2}})}$, where $\mu$ and $\delta$ are the mean and standard-deviation of normal distributions, respectively. As a choice of the quantum mechanical operator under consideration, we will assume the imaginary-time evolution generator corresponding to the free particle Hamiltonian; the kernel in the position-basis can then be written as $A^{G}_{ab}=\frac{1}{\sqrt{2\pi}\kappa}\exp{\Big(-\frac{1}{2\kappa^2}(x_{a}-x_{b})^2\Big)}$. Using this form of the operator kernel, it is possible to analytically compute the components of the matrix $\mathcal{L}_{ij}$, which is diagonal and of the form $\text{diag}(0, \frac{\kappa^4}{2\Delta^3})$, with $\Delta=2-\kappa^2\theta^{2}$, thus is independent of the mean $\mu$ of the distribution, a manifestation of the translation-invariance of the kernel. Thus in this special case, the rank-$1$ deformation only affects the $\theta^{2}$ direction of the parameter manifold; the curves of constant $\theta^{2}$ are the same as that of governed by the FIM.

%%%%%
\subsection{Quantum natural gradient with the loss-aware variational distance}
This loss-efficient pull-back metric now defines the distance and hence the effective learning rate on this variational manifold, and is expected to perform better than the standard natural gradient method based on the Fisher information metric only. Then the local solution of the optimisation problem for a small variation of the parameter $\delta\theta^{i}$ is the following
\begin{equation}
	g_{ij}^{\text{LA}}\delta\theta^{j}= -\eta\partial_{i}L(\theta)~,  \end{equation}
such that the optimal direction on the parameter manifold is 
\begin{equation}
	\theta_{t+1}^{i}=\theta^{i}_{t}-\eta g^{(\text{LA})ij}\partial_{j}L(\theta)~,
\label{LA-NG}
\end{equation}
provided that the metric is invertible.  This shows that for optimisation, the point $\theta^{i}_{t}$ must move in the \textit{opposite} direction to the natural gradient of the loss function with respect to $ g_{ij}^{\text{LA}}$. Substituting the Sherman-Morrison inverse metric formula for \eqref{LA-metric}, the LA-NG with respect to the LA-metric: eq. \eqref{LA-NG}, reduces to a QMT-NG problem with a modified step size, where, the effective step size for this LA-NG descent is 
\begin{equation}
	\eta^{\text{LA-NG}}_{\text{eff}}=\frac{\eta}{1+g^{(\text{FS})ij}\partial_{i}L\partial_{j}L},
\end{equation}
hence is always smaller than the natural gradient descent, where we have used the Einstein summation convention. Note that, due to non-trivial (pure) disformal transformation, the geodesic flow of the two metrics will not be similar, even though the local gradient flows are. We have provided a performance comparison of the QNG and the LA-QNG in section \ref{secperformance}, to explore the possible advantages of considering the LA modification.

%%%%%%%%%%%%%

\section{Conformal class of loss-aware metrics}\label{sec:CLAgeometry}
To consider the possibility of accelerating the loss-aware natural gradient, which  can be thought of as a proper change of the underlying geometry governed by the Fisher information metric, generated by a vector field (the derivatives of the loss functions), we introduce a novel class of metrics that not only ``stretches'' the geometry but also rescales  depending on a (loss-dependent) conformal factor. To this end, we consider the following class of metrics:
\begin{equation}
	g^{\text{CLA}}_{ij}(\theta)=\Omega^2(\theta)g_{ij}^{\text{LA}}=\Omega^2(\theta)\Big( g^{\text{FS}}_{ij}(\theta)+
		\partial_i L(\theta) \, \partial_j L(\theta)\Big),
\label{CLAmetric}
\end{equation}
for real and positive conformal factors $\Omega^2(\theta)$, which we will typically assume to be a function of the set of parameters $\{\theta\}$, induced by the loss. In the subsequent analysis, we  assume different forms of this conformal factor and will indicate these generic classes of metrics as conformal loss-aware (CLA) metrics.  Due to the inherent nature of the conformal transformations,  the direction of the gradient descent for both the LA-NG and CLA-NG is the same, since the conformal transformation scales every direction uniformly; though the exact amount of scaling may vary from point to point on the manifold. Similar to the pure rank-1 deformations, the distance measured by this conformally-modified metric is different from that of the distance measured with respect to the pure FIM case, however, unlike the previous case, this type of geometries changes (scales) the distances along the (images) of the level curves also.

Similar to the pure anisotropic case, the NG problem with the CLA-type metrics can also be written as a QNG problem, with the effective step-size on the other hand, is of the form 
\begin{equation}
    \eta^{\text{CLA-NG}}_{\text{eff}}=\frac{\eta}{\Omega^{2}(\theta)(1+g^{(\text{FS})ij}L\partial_{j}L)}=\Omega^{-2}(\theta)\eta^{\text{LA-NG}}_{\text{eff}},
\end{equation}
which indicates that the conformal factor effectively rescales the learning rate, and it is indeed possible to control the effective step size by tuning the overall conformal factor. To demonstrate the performance of CLA-type geometrics on the variational manifold, as well as the possible stability issues of the corresponding natural gradient descents near and far from the critical points, we will consider three types of CLA geometries, where the conformal factors have either exponential or power law dependence on (functions of) the associated loss functions.

\subsection{Case-1: CLA-$1$}
\label{CLA-1}
Let us first consider the following CLA geometries of the form \eqref{CLAmetric}, where the conformal factor is of the form  
\begin{equation}
    \Omega^2(\theta)=e^{C{(\theta)}},
\end{equation}
with $C{(\theta)}=\gamma \log(1+g^{(\text{FS})ij}\partial_{i}L\partial_{j}L)$ for an external control parameter $0<\gamma<1$ \footnote{In principle, we can of course assume the value of $\gamma$ to be $\gamma\geq 1$, which will still define a valid conformal modification; however, from an optimisation point of view, this choice might not always represent a situation with stable learning rates; see, in particular, the discussion for the CLA-$2$ geometries as well as the numerical implementation of the CLA geometries in subsection \ref{sec:convergence}.}. Then the effective learning rate for this type of conformal class of metrics parametrised by $\alpha$ is 
\begin{equation}
    \eta^{\text{CLA-$1$-NG}}_{\text{eff}}=\frac{\eta^{\text{LA-NG}}_{\text{eff}}}{(1+g^{(\text{FS})ij}\partial_{i}L\partial_{j}L)^{\gamma}},
\end{equation}
which reduces to $\eta^{\text{LA-NG}}_{\text{eff}}$ for the special case $\gamma=0$. The positive definiteness of  QMT and the nature of the parameter $\gamma$ ensure that the effective learning rate in this case is lower than that generated by LA-QMT. As can be seen, this form of conformal transformation stretches the geometry; distances between two points (along properly chosen curves) governed by this length functional are greater than those of both QMT and LA-QMT. 

%%%%
\subsection{Case-2: CLA-$2$}
As a second example, let us choose the following conformal factor, 
\begin{equation}
    \Omega^2(\theta)=e^{-C{(\theta)}},
\label{eqcfactor2}
\end{equation}
with $C{(\theta)}=\gamma\frac{g^{(\text{FS})ij}\partial_{i}L\partial_{j}L }{1+g^{(\text{FS})ij}\partial_{i}L\partial_{j}L}$, for the same choice of $\gamma$. The effective step-size for the natural gradient problem is then of the form  
\begin{equation}
    \eta^{\text{CLA-$2$-NG}}_{\text{eff}}=e^{C{(\theta)}}\eta^{\text{LA-NG}}_{\text{eff}}.
\label{CLA-2}
\end{equation}
This shows that the geometry determined by this conformal factor \eqref{CLA-2} has a learning rate greater than that of the LA-NG, on the other hand, since the conformal factor satisfies the inequality $1\leq e^{C{(\theta)}}\leq e^{\gamma}$, the effective step-size is bounded from above. Unlike the CLA-1, the choice of the conformal factor \eqref{eqcfactor2} `shrinks' the geometry; and the distance between two given points is less than all the other three geometries considered so far. 

\subsection{Case-3: CLA-$3$}
In order to increase the efficiency of the CLA-NG without strongly affecting the stability, in particular near the critical points, we next consider the following class of CLA metrics; where the conformal factor is of the form 
\begin{equation}
    \Omega^2(\theta)=(1+g^{(\text{FS})ij}\partial_{i}L\partial_{j}L)^{-\gamma},
\end{equation}
for the same range of the parameter $\gamma$. In this case, the effective step-size is of the form 
\begin{equation}
    \eta^{\text{CLA-$3$-NG}}_{\text{eff}}=(1+g^{(\text{FS})ij}\partial_{i}L\partial_{j}L)^{\gamma}\eta^{\text{LA-NG}}_{\text{eff}},
\end{equation}
from which we can obtain, using the Bernoulli inequality, the relation $ \eta^{\text{CLA-$3$-NG}}_{\text{eff}}\leq (1+\gamma \sigma)\eta^{\text{LA-NG}}_{\text{eff}}=\frac{1+\gamma\sigma}{1+\sigma}\eta$, where we have used the notation $\sigma=g^{(\text{FS})ij}\partial_{i}L\partial_{j}L$.
Even though the inverse of the conformal factor grows without bound as $\sigma\rightarrow{\infty}$, the effective learning is capped below the base learning rate $\eta$. In fact, the following inequality is satisfied $\eta^{\text{LA-NG}}_{\text{eff}}\leq  \eta^{\text{CLA-$3$-NG}}_{\text{eff}} \leq \eta$, and can achieve power-law interpolation depending on the value of $\gamma$. 

\begin{figure}
	\centering
\includegraphics[width=0.55\linewidth]{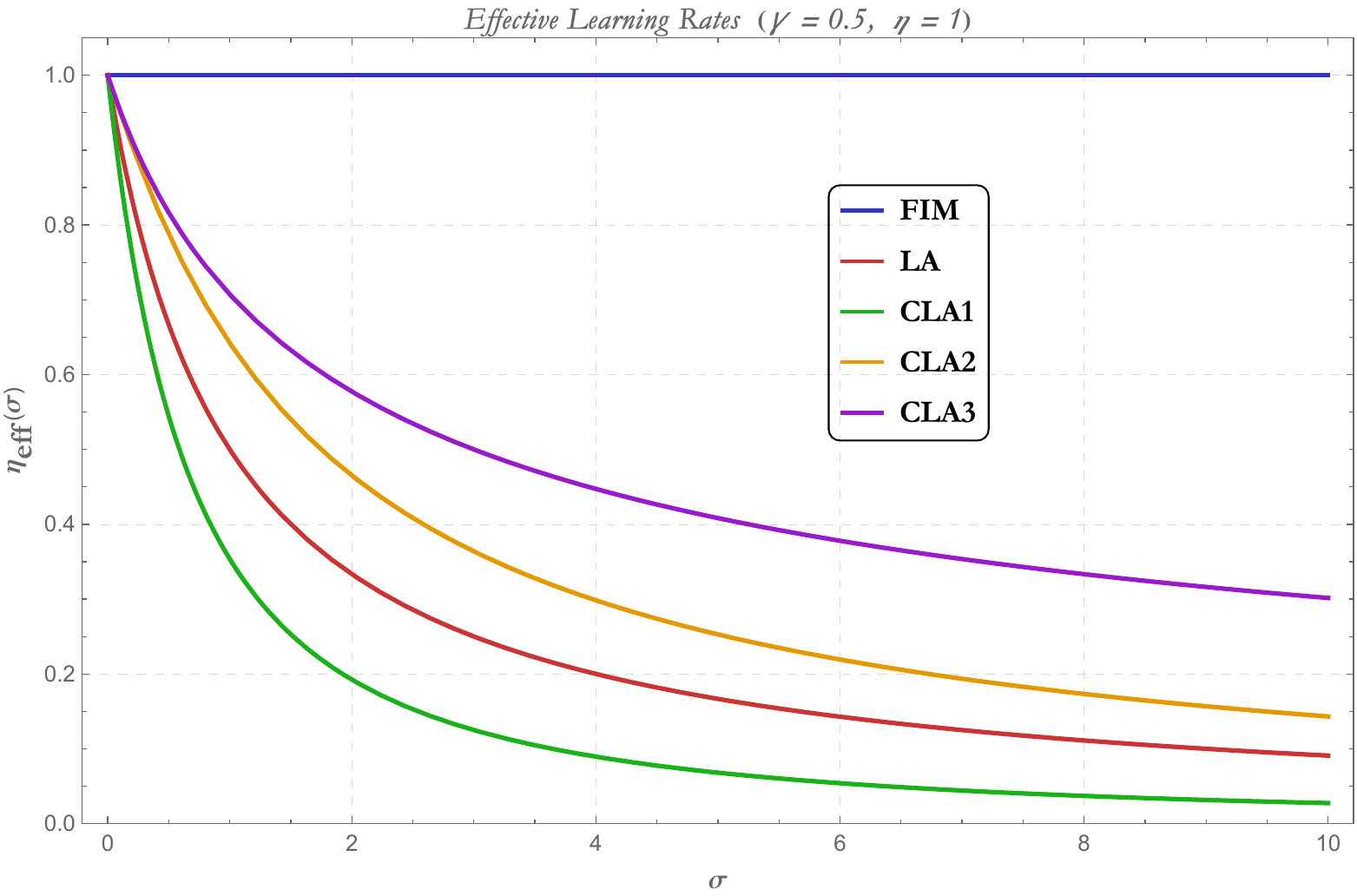}
	\caption{Comparison of the effective learning rates in different LA-metrics with the norm of the cost vector $\sigma$ with respect to the FIM. We have taken the base learning rate $\eta$ to be unity and the free parameter $\gamma=0.5$.}
	\label{fig:effective_rate}
\end{figure}

As a summary of the four modifications we have discussed so far; in fig.\ref{fig:effective_rate}, we have plotted the effective learning rates of the LA-modification of the geometries with the squared norm of the loss vector with respect to the FIM, $\sigma$, where we have assumed the base learning rate $\eta=1$. As can be seen, the CLA-$3$ modification has the highest learning rate among the four modifications presented; where, CLA-1 is the one most damped.

%%%%%%%%%%%%%%%%%%%

\section{Performance comparison of different loss-aware gradient descent techniques}\label{secperformance}

\subsection{Distributed hyperparameter Optimization of Quantum Natural Gradient Methods}
We implement a distributed hyperparameter optimisation to benchmark quantum natural gradient based optimisers in variational quantum circuits. We consider a system of $n=6$ qubits with Hilbert space dimension $\dim = 2^n$. The variational ansatz consists \cite{Stokes} of a circuit with $L=5$ layers and in total $n \times L = 30$ parameters. In our construction, each layer comprises single-qubit rotations followed by nearest-neighbour entangling gates; for further details, we refer the reader to \cite{McClean:2018jps}. The initial state we have considered is of the form
\begin{equation}
|\psi_0\rangle = |0\rangle^{\otimes n},
\end{equation}
which is preceded by a global rotation 
\begin{equation}
\textbf{U}_{\mathrm{init}} = \bigotimes_{i=1}^{n} R_y\!\left(\frac{\pi}{4}\right).
\end{equation}

Each layer applies parametrised rotations chosen from $\{R_x, R_y, R_z\}$, the standard Pauli operators; followed by 1D nearest-neighbour-controlled gates $Z$. The parametrised single-qubit rotations are chosen randomly for each circuit instance. At each layer $l$ and qubit $q$, the single-qubit rotation axis is chosen from the set $\{R_x, R_y, R_z\}$ with uniform probability.
Given a vector-valued parameter $\theta \in \mathbb{R}^{nL}$, the rotation applied in the layer $l$ and the qubit $q$ is
\begin{equation}
\mathbf{U}_{\mathrm{rot}}^{(l,q)}(\theta_{l,q}) =
\begin{cases}
R_x(\theta_{l,q}) & \text{if } c_{l,q} = 0, \\
R_y(\theta_{l,q}) & \text{if } c_{l,q} = 1, \\
R_z(\theta_{l,q}) & \text{if } c_{l,q} = 2.
\end{cases}
\end{equation}

The full rotation layer is given by the tensor product
\begin{equation}
\mathbf{U}_{\mathrm{rot}}^{(l)}(\theta_{l}) =
\bigotimes_{q=1}^{n}
\mathbf{U}_{\mathrm{rot}}^{(l,q)}(\theta_{l,q}).
\end{equation}

The parameters are initialised randomly according to a Gaussian distribution, $\theta_{l,q} \sim \mathcal{N}(0, 1)$ and subsequently rescaled as $\theta_{l,q} \rightarrow 2\pi \,\theta_{l,q}$. Thus, each circuit instance is specified by an independent draw of $\{\theta_{l,q}\}$ and $\{c_{l,q}\}$ as 
\begin{equation}
\mathbf{U}(\theta) =
\prod_{l=1}^{L}
\left(
\mathbf{U}_{\mathrm{ent}}^{(l)} \, \mathbf{U}_{\mathrm{rot}}^{(l)}(\theta)
\right)
\mathbf{U}_{\mathrm{init}}.
\end{equation}

The Hamiltonian is constructed from a randomly drawn $2$-qubit Hamiltonian $\mathbf{H}_2$, embedded into the full n-qubit system as
\begin{equation}
\mathbf{H} = \mathbf{H}_2 \otimes \mathbf{I}^{\otimes (n-2)},
\end{equation}
with a fixed spectral gap of $ \Delta E = E_{1} - E_{0} =1.5$. 

\subsection{Block-diagonal quantum geometric tensor}

We employ a block-diagonal approximation of the quantum geometric tensor (QGT) \cite{Stokes}, where each block corresponds to a circuit layer. For a given layer $l$, the covariance matrix is given by
\begin{equation}
gb^{(\mathrm{FS}), (l)}_{ij}
=
\frac{1}{4}
\left(
\langle P_i P_j \rangle
-
\langle P_i \rangle \langle P_j \rangle
\right),
\end{equation}
where $P_i$ are local Pauli operators in the $l^{th}$  layer determined by the circuit structure. To model imperfections in the QGT block QGT \cite{Koczor2022,Cerezo}, we introduce a simple multiplicative symmetric noise at each layer that has the form:
\begin{equation}
gb^{(\mathrm{FS})}_{ij}
\;\rightarrow\;
gb^{(\mathrm{FS})}_{ij}
+
\left(
\bigl| gb^{(\mathrm{FS})}_{ij} \bigr| 
+ \varepsilon \right)
\varsigma \, \beta_{ij},
\end{equation}
where $\beta_{ij} \sim \mathcal{N}(0,1)$ and the noise is symmetrised to preserve Hermiticity. The noise strength is fixed to $\varsigma = 0.1$. The $\varepsilon$ is a small positive regularisation constant introduced to ensure that the noise remains finite even when $g^{(\mathrm{FS})}_{ij}$ is close to zero. The optimisation schemes considered in this section are summarised in table~\ref{tab:optimizersqn}.

\begin{table}[t]
\centering
\renewcommand{\arraystretch}{1.4}
\begin{tabular}{p{6.5cm} p{4.9cm} p{5.0cm}}
%\begin{tabular}{c c}
\hline\hline
\textbf{Optimiser} & \textbf{Update rule} \\
\hline
Quantum Natural Gradient (QNG) &
$\displaystyle
\Delta\theta = -\left(gb \right)^{-1}\nabla E
$
\\[0.8em]
\hline
\\[0.3em]

loss-aware QNG (LA-QNG) &
$\displaystyle
\Delta\theta = -\left(gb + \xi \nabla E \nabla E^{\top} \right)^{-1}\nabla E
$
\\[1em]
\hline
\\[0.3em]
Conformal Loss-Aware QNG-2 
(CLA-$2$-QNG) &
$\displaystyle
\begin{aligned}
\sigma &= \xi\, \nabla E^{\top} gb \nabla E, \\
\Omega^2 &= \exp\!\left(-\frac{\gamma \sigma}{1+\sigma}\right), \\
\Delta\theta
&= -\left(\Omega^2 \left(gb + \xi \nabla E \nabla E^{\top}\right) \right)^{-1}\nabla E
\end{aligned}
$
\\[1.2em]
\\
\hline
\\[0.3em]
Conformal Loss-Aware QNG-3  (CLA-$3$-QNG) &
$\displaystyle
\begin{aligned}
\Omega^2 &= (1+\sigma)^{-\gamma}, \\
\Delta\theta
&= -\left(\Omega^2 \left(gb + \xi \nabla E \nabla E^{\top}\right) \right)^{-1}\nabla E
\end{aligned}
$
\\
\\
\hline\hline
\end{tabular}

\caption{Optimization algorithms compared in this work. Here $\xi$ and $\gamma$ are tunable hyperparameters and $\nabla E$ denotes the gradient of the energy with respect to the variational parameters, $\nabla E \equiv \frac{\partial E(\theta)}{\partial \theta}$, where $E(\theta) = \langle \psi(\theta) | \mathbf{H} | \psi(\theta) \rangle$.}
\label{tab:optimizersqn}
\end{table}

\subsection{Convergence and hyperparameter optimization}\label{sec:convergence}
In this subsection, we will describe the details of the convergence criteria and the hyperparameter optimisation used later on. 
In the LA-NG, both with the FIM and QMT as the parameter space metric, have two hyperparameters to optimise, $\xi$ and $\eta$, where in the conformal-scaled versions, we have another one, namely $\gamma$. To this end, we have taken each optimisation run to  proceed up to a maximum of $12000$ iterations and subsequently, the convergence is defined via the relative error
\begin{equation}
\frac{|E(\theta) - E_{\mathrm{exact}}|}{|E_{\mathrm{exact}}|} < 10^{-11},
\end{equation}
which must be satisfied for $400$ consecutive checks to ensure stability. We evaluate performance over an ensemble of $50$ random circuits, with initial parameters sampled as $\theta \sim \mathcal{N}(0,1)$ and random Pauli rotation axes chosen independently for each layer. Performance is quantified by the number of iterations required to reach convergence, and we report a $20\%$ trimmed mean to mitigate the effect of outliers. Hyperparameters are optimised using a distributed Bayesian optimisation procedure \cite{Shahriari2016TakingTH,Bergstra2011AlgorithmsFH}, implemented via the \texttt{Optuna} framework \cite{Akiba2019optuna} with the search space being
\begin{equation}
\mathrm{\eta} \in [10^{-3},10^{-1}], \quad
\xi \in [10^{-3},10^{-1}], \quad
\gamma \in [0,4].
\end{equation}

\begin{figure}[t]
\centering
\includegraphics[width=0.99\linewidth]{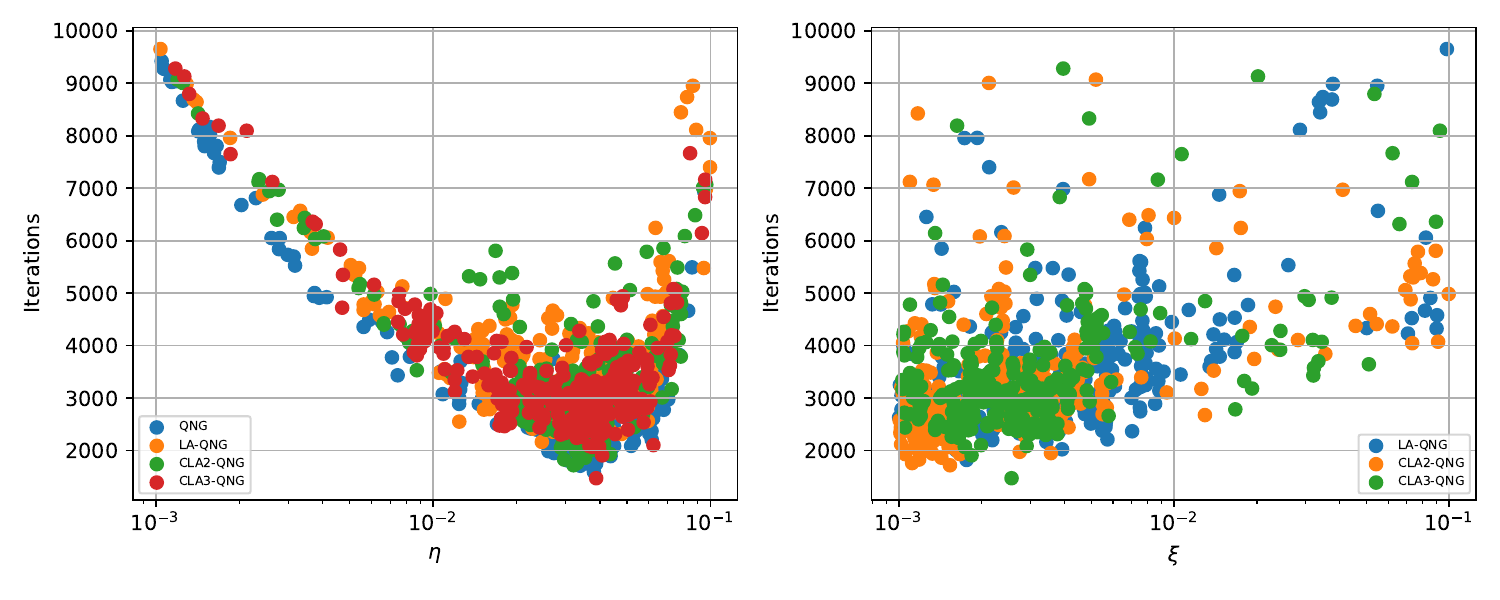}
\includegraphics[width=0.99\linewidth]{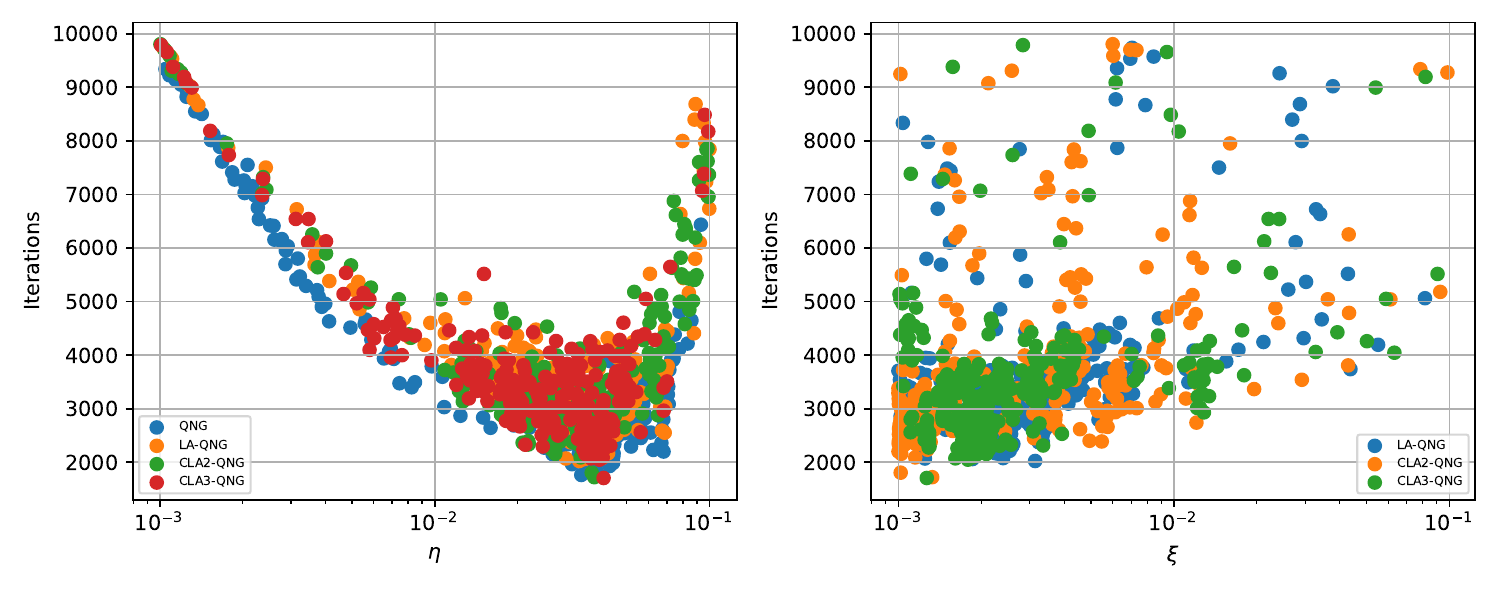}
\caption{
Learning landscapes of the optimization dynamics for $\varsigma = 0.1$ noise level (above) and without noise $\varsigma = 0$ (below).
(a) Iterations to convergence as a function of the learning rate $\eta$ for all optimizers.
(b) Dependence on $\xi$ for loss-aware methods.
}
\label{fig:learning_landscape1}
\end{figure}

\begin{figure}[t]
\centering
\includegraphics[width=0.49\linewidth]{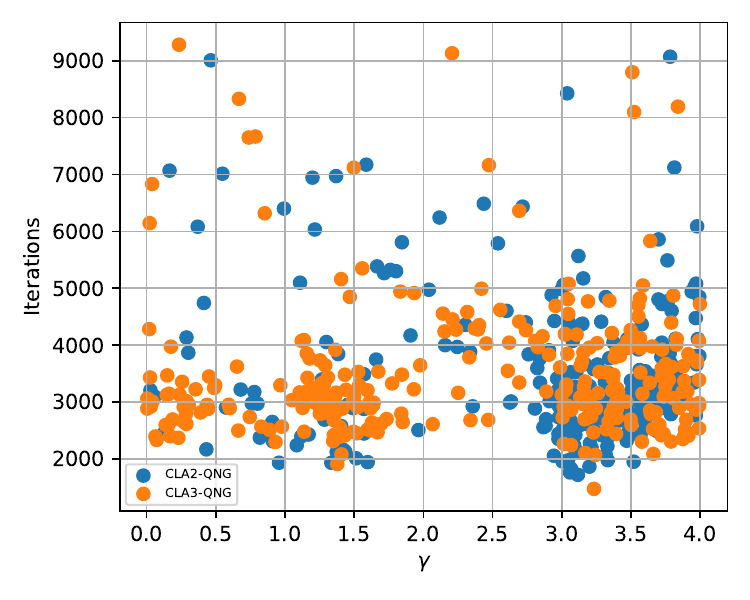}
\includegraphics[width=0.49\linewidth]{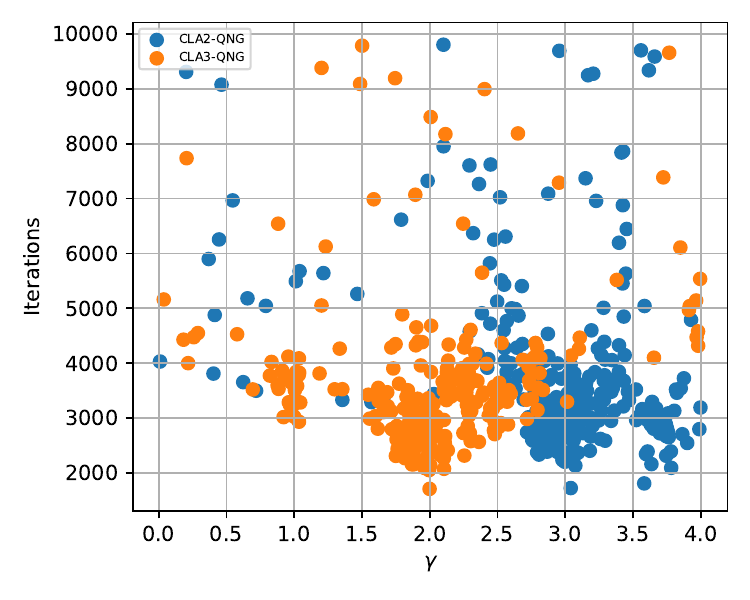}
\caption{
Learning landscapes of the optimization dynamics for $\varsigma = 0.1$ noise level (left) and without noise $\varsigma = 0$ (right).
}
\label{fig:learning_landscape2}
\end{figure}

From Figures~\ref{fig:learning_landscape1} and~\ref{fig:learning_landscape2}, we extract the optimal hyperparameter regimes for the different gradient-descent schemes. In the noisy case, all four methods exhibit a consistent optimal learning rate in the range $\eta \in [0.031,\,0.039]$, indicating a scheme-independent scale for stable convergence. Within this regime, the number of iterations decreases as $\eta$ increases, whereas for $\eta \gtrsim 0.045$ the iterations increase sharply, signalling the onset of instability. The optimal values of the curvature-related parameter $\xi$ are found to be $\xi \approx 1.5 \times 10^{-3}$ for LA-QNG, $\xi \approx 1.75 \times 10^{-3}$ for CLA-$2$-QNG, and $\xi \approx 2.56 \times 10^{-3}$ for CLA-$3$-QNG, showing a slight shift towards larger $\xi$ in the noisy setting. The optimal conformal scaling exponent is $\gamma \approx 3.116$ for CLA-$2$-QNG and $\gamma \approx 3.23$ for CLA-$3$-QNG, suggesting that strong damping ($\gamma \sim 3$) is preferred to stabilise the optimisation dynamics.

For the noiseless case, the optimal curvature-related parameter is $\xi \approx 3.14 \times 10^{-3}$ for DQNG, $\xi \approx 1.33 \times 10^{-3}$ for CL-DQNG, and $\xi \approx 1.27 \times 10^{-3}$ for CL2-DQNG, indicating a shift toward smaller $\xi$-values in the conformally scaled variants compared to standard DQNG. The optimal conformal scaling exponent is $\gamma \approx 3.04$ for CL-DQNG and $\gamma \approx 2.00$ for CL2-DQNG.

\subsection{Performance Comparison}

\begin{figure}[t]
\centering

\includegraphics[width=0.48\linewidth]{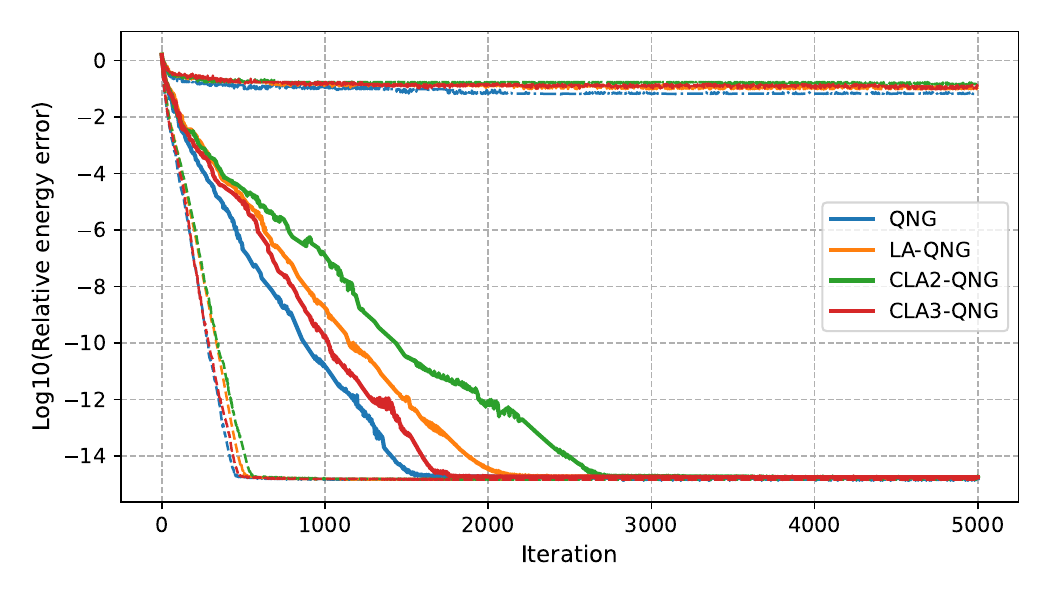}
\hfill
\includegraphics[width=0.48\linewidth]{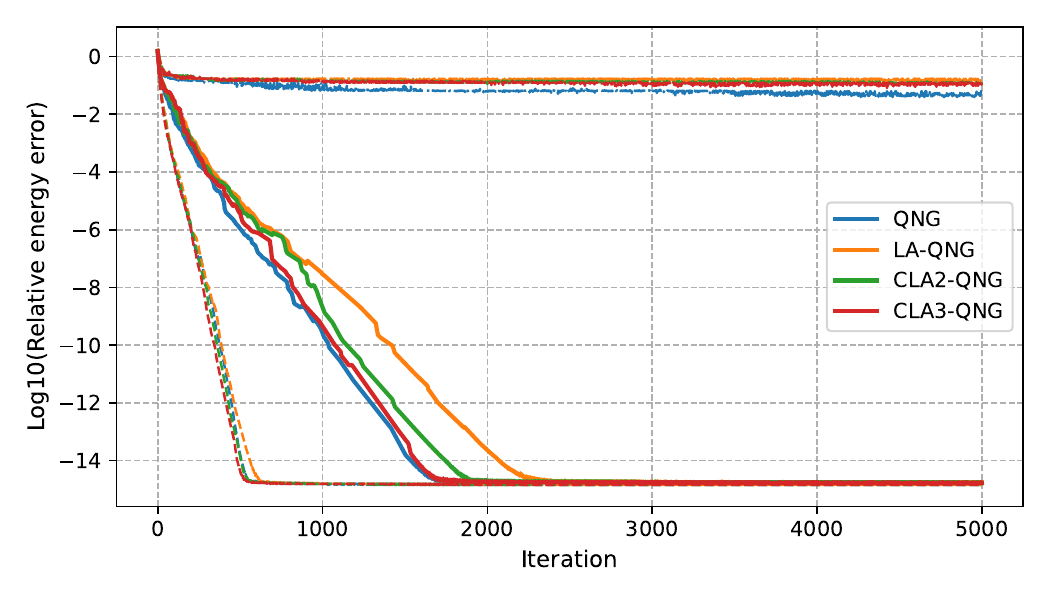}

\includegraphics[width=0.48\linewidth]{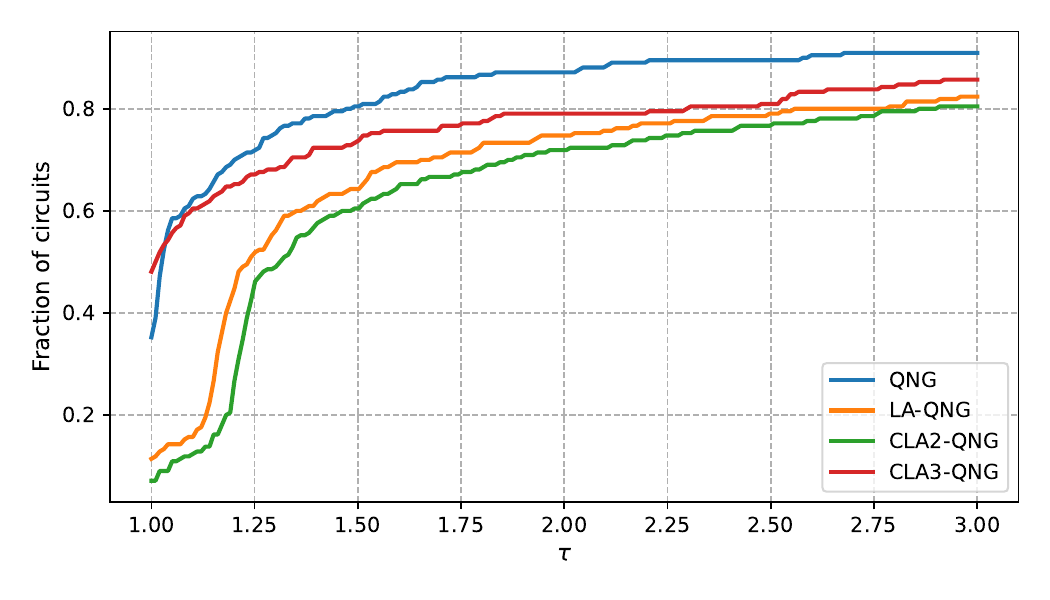}
\hfill
\includegraphics[width=0.48\linewidth]{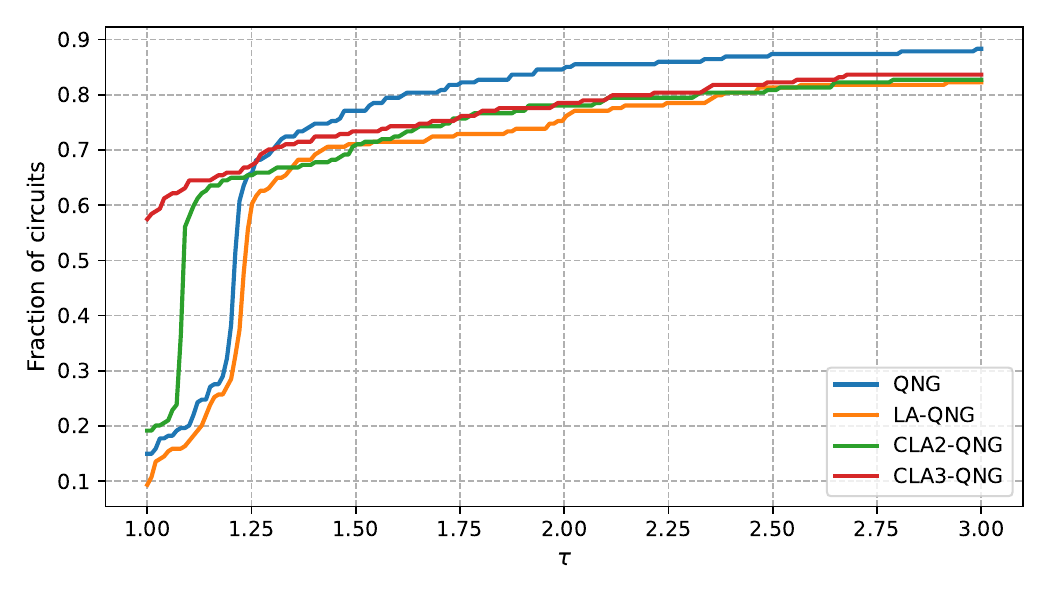}

\caption{
(a) Convergence of the median relative energy error as a function of iteration (Above), with dashed lines indicating the interquartile range across circuit instances for $\varsigma = 0.1$ (left) and $\varsigma = 0$ (right). 
(b) Performance profile (Dolan--Mor\'e) showing the fraction of circuits (Below) solved within a factor $\tau$ of the best method. 
}
\label{fig:optimization_results}
\end{figure}

Figure~\ref{fig:optimization_results} summarises the convergence behaviour across different optimisation schemes, all results are obtained over an ensemble of $300$ circuits to avoid finite-sample effects, using the optimal hyperparameters determined from the preceding optimisation procedure. We observe that the median convergence rate is fastest for QNG, followed by CLA-$3$-QNG, then LA-QNG, with CLA-$2$-QNG performing the slowest overall. The dashed curves represent variability across circuit instances, where the lower dashed curve corresponds to the $25^\mathrm{th}$ percentile and the upper dashed curve to the $75^\mathrm{th}$ percentile. Notably, in the top $25\%$ percentile (fastest runs), CLA-$3$-QNG exhibits a convergence behaviour comparable to QNG, indicating that its best-case performance matches that of the standard natural gradient method. This highlights that while QNG is optimal on average, CLA-$3$-QNG remains competitive in favourable regimes.

\begin{figure}[t]
\centering
\includegraphics[width=0.49\linewidth]{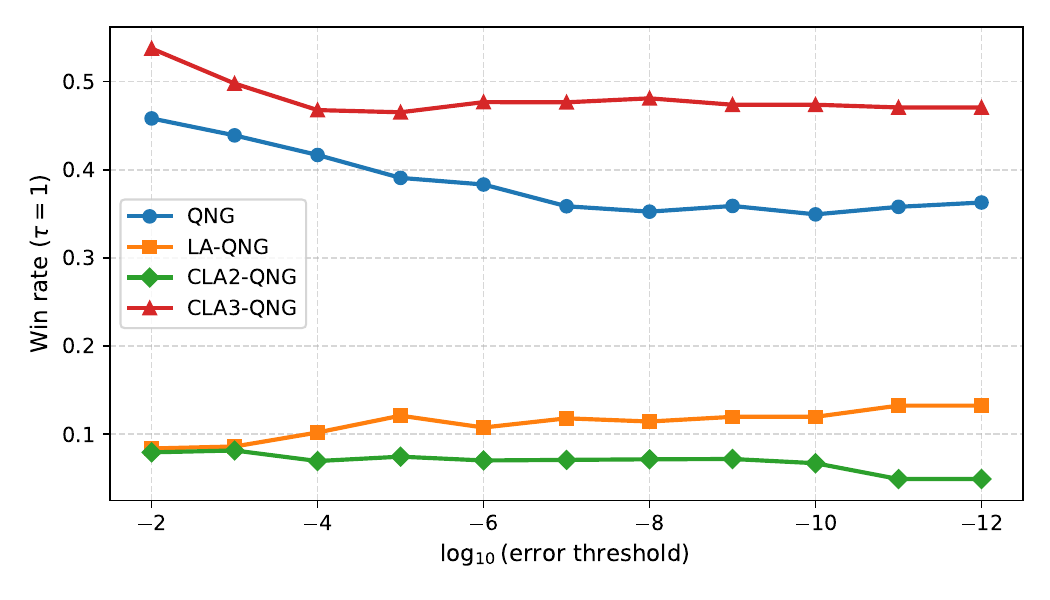}
\includegraphics[width=0.49\linewidth]{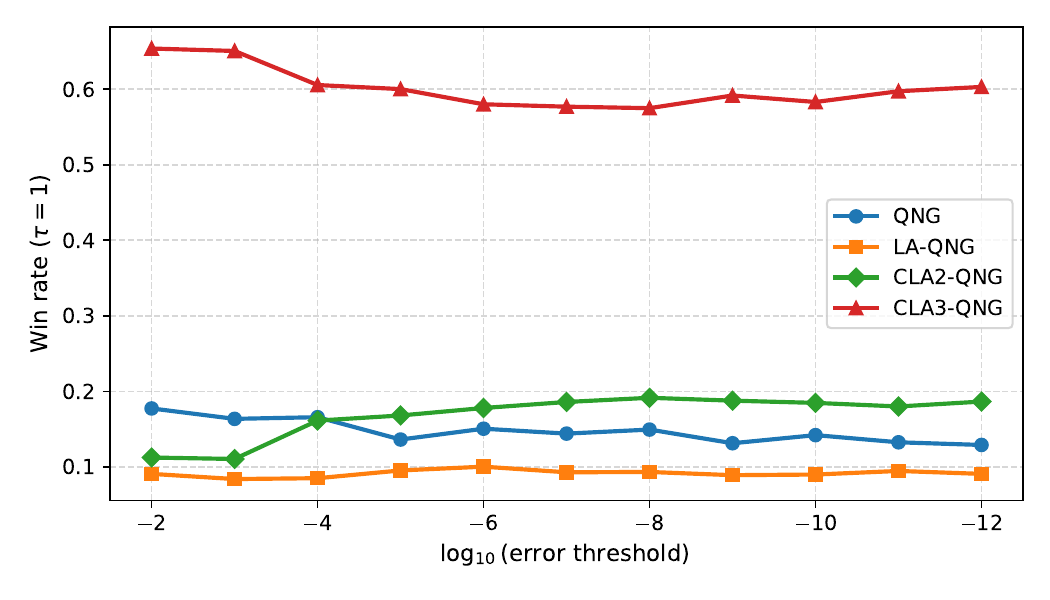}
\caption{
Win rate ($\tau = 1$) as a function of the error threshold. The win rate measures the fraction of circuits for which a given method achieves the best convergence time for $\varsigma =0.1$ (left) and $\varsigma = 0$ (right). As the threshold is varied, this plot reveals the relative robustness of different optimization schemes across accuracy regimes.
}
\label{fig:winrate}
\end{figure}

The Dolan--Mor\'e \cite{2001cs........2001D,PhysRevResearch.2.043246,Arrasmith:2020kvg} performance profile figure~\ref{fig:optimization_results} on the benchmark is restricted to the subset of solved instances, where $210$ out of $300$ circuits satisfy the convergence criterion for at least one optimisation method. The Dolan--Mor\'e performance profile is constructed using a convergence threshold of $10^{-8}$ for the relative error. For each circuit, the performance ratio is defined as $\tau = t / t_{\mathrm{best}}$, where $t$ is the iteration count of a given method and $t_{\mathrm{best}}$ is the best achieved among all methods for that instance. The plotted curves represent the fraction of circuits solved within a factor $\tau$ of the best method. At $\tau = 1$, CLA-$3$-QNG achieves the highest value, indicating that it most frequently achieves the absolute best convergence time across instances. However, as $\tau$ increases, QNG rapidly overtakes and dominates the profile, achieving a higher fraction of near-optimal performance across circuits. In the noiseless case, CLA-$2$-QNG is the second-best performer near $\tau = 1$, while QNG becomes the dominant method by $\tau \approx 1.25$. This indicates that while CLA-$3$-QNG can outperform other methods in favourable instances, its performance is less consistent, whereas QNG provides the most robust and reliable convergence overall.

Figure~\ref{fig:winrate} shows the dependence of the win rate ($\tau = 1$) on the error threshold. For the noisy case, we observe that CLA-$3$-QNG consistently achieves the highest win rate across the range $10^{-2}$ to $10^{-12}$, winning approximately $50\%$ of the circuits. In comparison, QNG attains a win rate of about $40\%$ over the same range. In the noiseless setting, this trend is further amplified, with CLA-$3$-QNG achieving a win rate of around $60\%$, while QNG drops to approximately $20\%$. This indicates that while QNG is more robust on average, CLA-$3$-QNG more frequently achieves the absolute best convergence time, highlighting its advantage in favourable instances. These results highlight a fundamental trade-off between robustness and peak performance, with QNG ensuring consistent convergence across circuits, while CLA-$3$-QNG excels in achieving the fastest solutions in favourable instances.

\subsection{Classical Optimization with Fisher-Based Preconditioning}

We study optimiser behaviour on MNIST using a simple multilayer perceptron (MLP) with two hidden layers of width 64 and GELU activations, followed by a linear output layer for 10-way classification. The input images are flattened to 784-dimensional vectors and normalised to $[0,1]$. Both training and evaluation are performed using shuffled minibatches of size $1024$, with accuracy of the full-dataset  computed by aggregating predictions over all minibatches at validation intervals \cite{Harvey2025optimiser}. 
\begin{table}[t]
\centering
\renewcommand{\arraystretch}{1.35}
\setlength{\tabcolsep}{4pt}
\begin{tabular}{p{3.0cm} p{4.9cm} p{5.0cm}}
\hline
\textbf{Optimizer} & \textbf{Update rule} & \textbf{Variables} \\
\hline
\\
SGD-RMS &
\(\begin{aligned}
\Delta\theta_t
&= -\eta\,\gamma_t\,
\frac{m_t}{(1-\beta_m^t)(\sqrt{\hat r_t}+\varepsilon)}
-\eta\lambda\theta_t
\end{aligned}\)
&
\(\begin{aligned}
d_t &= \nabla_\theta L(\theta_t),\\
r_t &= \beta_{\mathrm{rms}} r_{t-1} + (1-\beta_{\mathrm{rms}}) d_t^2, \quad
\hat r_t = \frac{r_t}{1-\beta_{\mathrm{rms}}^t},\\
m_t &= \beta_m m_{t-1} + (1-\beta_m) d_t,\\
\mu_t &= \beta \mu_{t-1} + (1-\beta)\,\xi \sum_i
\frac{d_{t,i}^2}{\sqrt{\hat r_{t,i}}+\varepsilon}, \quad
\hat\mu_t = \frac{\mu_t}{1-\beta^t},\\
\gamma_t &= \frac{1}{1+|\hat\mu_t|}.\\
\end{aligned}\)
\\
\\
\hline
\\
F-NG &
\(\Delta\theta_t = -\eta\,p_t\)
&
\(\begin{aligned}
d_t &= \nabla_\theta \tilde{L}(\theta_t),\\
m_t &= \beta_m m_{t-1} + (1-\beta_m)d_t,\\
p_t &= P_t^{-1}m_t.\\
\end{aligned}\)
\\
\\
\hline
\\
F-LANG &
\(\Delta\theta_t = -\eta\,\frac{1}{1+s_t}\,p_t\)
&
\(\begin{aligned}
d_t &= \nabla_\theta \tilde{L}(\theta_t),\\
m_t &= \beta_m m_{t-1} + (1-\beta_m)d_t,\\
p_t &= P_t^{-1}m_t,\\
s_t &= \langle m_t,p_t\rangle.\\
\end{aligned}\)
\\
\\
\hline
\\
F-CLA-$2$ &
\(\Delta\theta_t
= -\eta\,
\frac{\exp\!\left(\gamma \frac{s_t}{1+s_t}\right)}{1+s_t}\,p_t\)
&
\(\begin{aligned}
d_t &= \nabla_\theta \tilde{L}(\theta_t),\\
m_t &= \beta_m m_{t-1} + (1-\beta_m)d_t,\\
p_t &= P_t^{-1}m_t,\\
s_t &= \langle m_t,p_t\rangle.\\
\end{aligned}\)
\\
\\
\hline
\\
F-CLA-$3$ &
\(\Delta\theta_t = -\eta\,(1+s_t)^{\gamma-1}\,p_t\)
&
\(\begin{aligned}
d_t &= \nabla_\theta \tilde{L}(\theta_t),\\
m_t &= \beta_m m_{t-1} + (1-\beta_m)d_t,\\
p_t &= P_t^{-1}m_t,\\
s_t &= \langle m_t,p_t\rangle.\\
\end{aligned}\)
\\
\\
\hline
\end{tabular}

\vspace{2mm}
\parbox{0.96\linewidth}{\small
}

\caption{Summary of the optimiser updates used in this work.}
\label{tab:optimizers}
\end{table}

We benchmark standard first-order optimisers against curvature-aware methods discussed in this work. For completeness, we have summarised the LA-geometries and the corresponding update rules in table \ref{tab:optimizers}. As first-order baselines, we employ Adam and SGD-RMS \cite{Harvey2025optimiser}, both of which rely solely on gradient-based updates without incorporating curvature information. The remaining optimisers F-NG , F-LANG, F-CLA-$2$ and F-CLA-$3$ are custom implementations that leverage K-FAC to approximate the FIM and precondition the update direction accordingly. These methods differ in the scalar rescaling applied to the preconditioned gradient, enabling a systematic study of Fisher-informed optimisation dynamics. Adam and SGD-RMS therefore serve as clean baselines for isolating the effect of curvature information.

In particular, the curvature matrix is approximated as \(P_t \approx F_t + \tilde{\delta} I\), where \(\tilde{\delta}\) is a damping parameter that stabilises the inversion. The preconditioned direction \(p_t = P_t^{-1} m_t\) therefore implicitly incorporates damping introduced, with \(P_t^{-1}\) denoting the K-FAC approximation to the inverse FIM \cite{Martens2015,Grosse2016,Martens2014,Dangel2025}. Additionally, regularisation with coefficient \(\lambda\) is included in the loss function,
\begin{equation}
   \tilde{L}(\theta) = L(\theta) + \frac{\lambda}{2}\|\theta\|^2, 
\end{equation}
and thus enters through the gradient and curvature estimates rather than appearing explicitly in the update equations. For K-FAC, curvature is tracked via an exponential moving average with coefficient $\beta_{\mathrm{curv}}$, updated every $20$ steps, with inverse updates every $40$ steps.

\begin{figure}[t]
\centering
\includegraphics[width=0.48\linewidth]{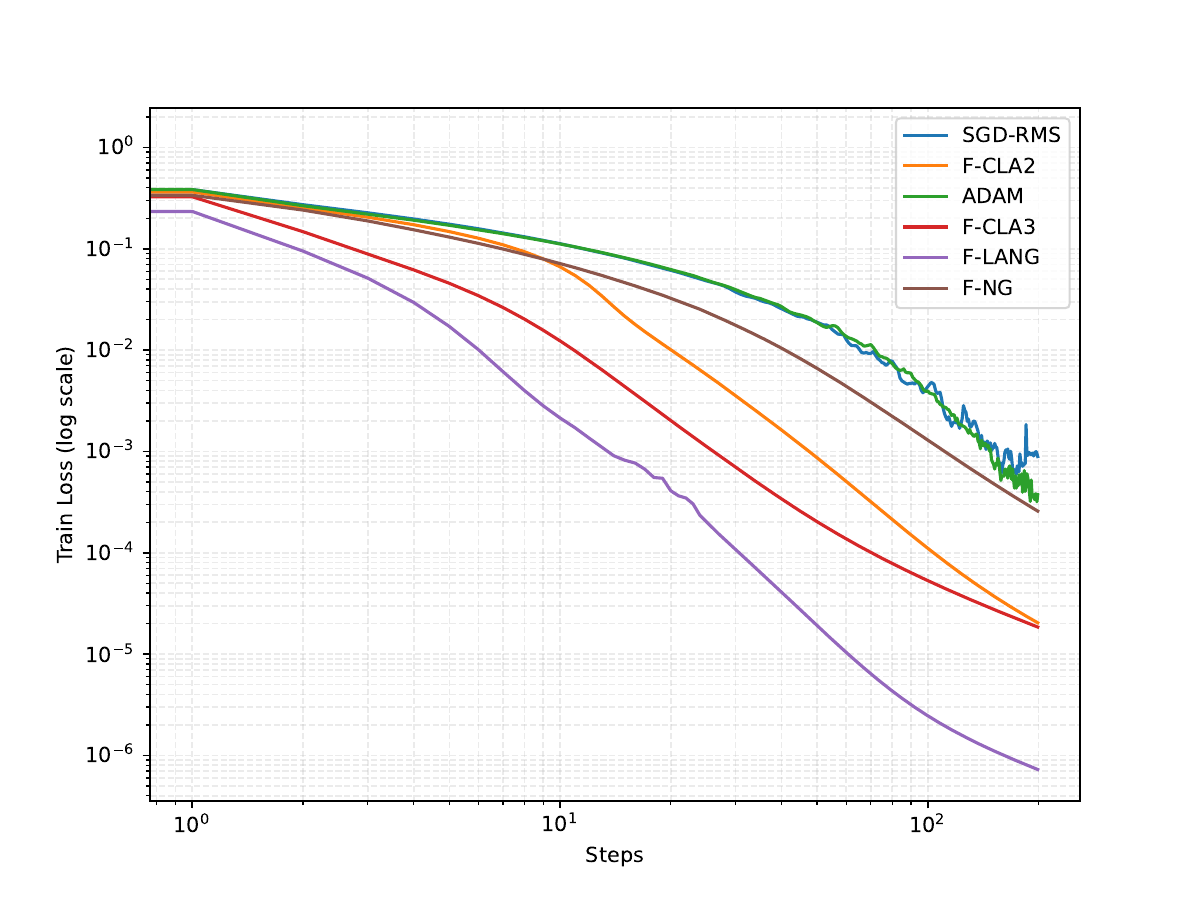}
\includegraphics[width=0.44\linewidth]{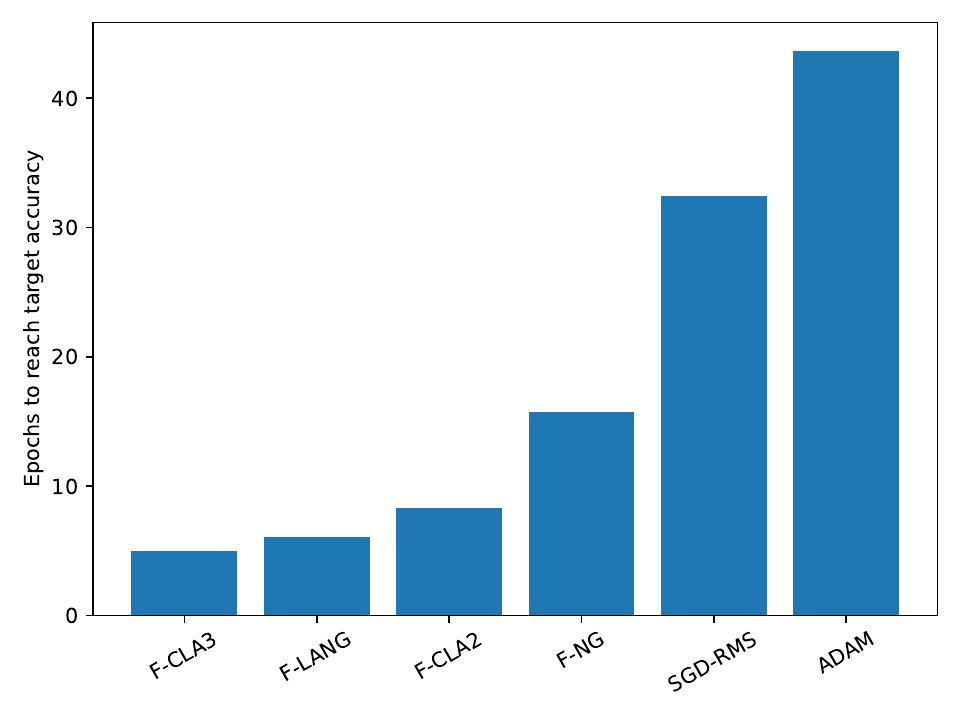}
\caption{Training loss (left) as a function of optimisation steps. Curves show the mean over the top-$15$ runs for each optimiser. Time-to-threshold (right) for different optimisers, defined as the number of epochs required to reach a fixed validation accuracy of $0.973$.}
\label{fig:cl_train_loss}
\end{figure}

We analyse the training dynamics by aggregating multiple independent runs for each optimiser, generated from 100 hyperparameter configurations explored using Bayesian optimisation \cite{Shahriari2016TakingTH}, where, to ensure the consistency and comparability, we impose a strict validation criterion on the runs. We have considered a run to be valid only if it contains a complete training trajectory of exactly 200 steps, with no missing values in the training loss. Runs that fail any of these conditions are subsequently discarded. For each optimiser, we rank valid runs based on their minimum training throughout the run. The top \(15\) runs with the lowest training loss are selected. This ensures that only the best-performing and fully converged trajectories are retained for analysis. We plot in Fig.~\ref{fig:cl_train_loss} the mean training loss as a function of optimisation steps for each optimiser. Curvature-aware methods that incorporate K-FAC preconditioning achieve significantly lower training loss compared to Adam and SGD-RMS. Figure~\ref{fig:cl_train_loss} shows the time-to-threshold for each optimiser, defined as the number of epochs required to achieve a fixed validation accuracy. The K-FAC-based methods achieve significantly lower values, indicating faster convergence. In contrast, Adam and SGD-RMS require substantially more epochs and exhibit comparable performance, highlighting the advantage of curvature-aware optimisation. F-CLA-$3$ achieves the best performance, followed by F-LANG, both significantly outperforming the other methods.

%%%%%%%%%%%%%
\section{Loss-aware parameter space geometry from a parameter-space overlap of states}\label{secdivergence}
So far, we have discussed how the embedding of the loss function in the parameter manifold leads to the `natural' emergence of the LA-geometry in terms of the pull-back-geometry on the loss hypersurface. This, however, is one of the several different ways to introduce the inherent geometry of the observable-space into devising efficient optimisation algorithms. To be more specific, a first-principle information geometric construction of induced geometry on the statistical manifold with the knowledge of the parameter dependence of the observable space should start from a well-defined divergence function that has information about the parameter sensitivity of both the space of PDFs as well as the observables. As a reminder, at this point we note that the FIM can be obtained as a unique expansion of the divergence functions that have support on the space of parametrised PDFs \cite{Amari2016information, Amari2000methods}. 
This motivates us to explore an analogue construction specifically for quantum systems based  on an overlap function having support on the corresponding states, which are vectors on the projective Hilbert spaces, and importantly also on the space of (parametrised) observables, which  are linear, Hermitian operators acting on the Hilbert space. Even though the geometry of the state space of quantum systems has a long and rich history, such a  construction that takes into account both the state space as well as the parametric dependence of the possible outcomes on that Hilbert space, to the best of our knowledge, has not been worked out before.  In this section, we aim to outline such a formalism, where, first, we will construct the overlap function of perturbed states, then provide a gauge-invariant construction of the resulting geometries. Then we will generalise this method to two mutually biorthogonal functions, which is essential for obtaining an analogue notion of dual $\pm\alpha$-connections. 

\subsection{Construction and implementing the gauge-invariance}
The Hermitian inner product on the Hilbert space of quantum states naturally induces a Riemannian metric on the projective Hilbert space of such states, which essentially is the physically relevant gauge-invariant sector of the full Hilbert space. This in turn, when pulled-back to the parameter (sub) manifold of interest gives a useful notion of distance between two nearby quantum states; which can be obtained from considering the following overlap \cite{Provost1980riemannian}
\begin{equation}
\mathcal{D}\Big(\Psi(\theta+\delta\theta),\Psi(\theta)\Big)=\braket{\Psi(\theta+\delta\theta)-\Psi(\theta)|\Psi(\theta+\delta\theta)-\Psi(\theta)}.
\label{PVoverlap}
\end{equation}
The expansion of this overlap for a properly normalised initial state $\braket{\Psi(\theta)|\Psi(\theta)}=1$, and the retention of terms up to the second order, we can obtain the following form of the quantum geometric tensor (QGT) 
\begin{equation} 
{FS}_{ij}=\braket{\partial_{i}{\Psi(\theta)}|\partial_{j}\Psi(\theta)}-\braket{\partial_{i}\Psi(\theta)|\Psi(\theta)}\braket{\Psi(\theta)|\partial_{j}\Psi(\theta)}~,
\label{tensorstructure}
\end{equation}
where we have to subtract a projection to make the QGT physically meaningful. In a similar motivation, here we consider how the change of the cost function with respect to a small change of the parameter $\delta\theta$ affects LA-QMT. To this end to incorporate both the geometry of the parameter manifold and the loss landscape, we propose to study the following overlap 

\begin{equation}
\mathcal{D}_{LA}\Big(\Psi(\theta+\delta\theta),\Psi(\theta)\Big)=\braket{\Psi(\theta+\delta\theta)-\Psi(\theta)|(\mathbf{I}+\frac{\mathbf{A}}{L(\theta)})|\Psi(\theta+\delta\theta)-\Psi(\theta)},
\label{PVoverlap}
\end{equation}

where $\mathbf{I}$ is the identity operator on the Hilbert space. Expanding this overlap and considering second-order terms, we will recover the standard Fubini-Study tensor (of course, without the gauge fixing term) and the ``response'' of the operator $\mathbf{A}$ with respect to the small change in the parameter, which is of the form ${FS}^{\mathbf{A}}_{ij}=\braket{\partial_{i}{\Psi(\theta)}|\mathbf{A}|\partial_{j}\Psi(\theta)}$. In this sense, we can consider the  FS tensor (gauge-dependent) as a response of the identity operator $\mathbf{I}$ for a small change of the state parameter. Although this term ${FS}^{\mathbf{A}}_{ij}$ transforms as a rank-$2$ tensor under a change of coordinates, it is still not meaningful to consider it as a LA-QGT, since it is not invariant under a (possibly complicated) parameter-dependent phase transformation. In the subsequent analysis, we assume that the operator $\mathbf{A}$ is Hermitian with respect to the canonical inner product on the Hilbert space. Using the same identification of the transformation of  $\ket{\partial_{i}{\Psi(\theta)}}$ under a gauge transformation, as was used in \cite{Provost1980riemannian}, we can get a gauge-invariant form of ${FS}^{\mathbf{A}}_{ij}$, which do qualify as a proper LA-QGT, of the form 
\begin{equation} 
\begin{split}
{FS}^{\text{LA}}_{ij}=FS_{ij}+\frac{1}{L}\braket{\partial_{i}{\Psi(\theta)}|\mathbf{A}|\partial_{j}\Psi(\theta)}-\frac{1}{L^2}\braket{\partial_{i}\Psi(\theta)|\mathbf{A}|\Psi(\theta)}\braket{\Psi(\theta)|\mathbf{A}|\partial_{j}\Psi(\theta)}.
\label{LAtensorstructure}
\end{split}
\end{equation}
This Hermitian tensor \eqref{LAtensorstructure} can be decomposed into two parts, similar to the usual FS tensor, which are the real as well as the symmetric part \textit{and} the purely imaginary as well as the anti-symmetric part. This indicates how the geometry of the loss landscape can possibly alter not only the Riemannian metric on the parameter manifold but  also can have a non-trivial impact on the Berry curvature (together they define the symplectic structure on the Projective Hilbert space in the standard case) as well. Note that this is in contrast with the construction of LA-QMT in \ref{LA-QMT-1}, where by construction the Berry curvature remains unchanged \footnote{Of course, this is because the construction in section \ref{LA-QMT-1} is motivated by the modification of the classical FIM, and hence only concentrates on the specific real part of the full FS tensor in the quantum case.}. However, it should be noted that the geometry induced by the tensor structure \eqref{LAtensorstructure} is an operator-weighted geometry that is not on a similar footing with the projective geometry of pure state-space \footnote{It might be possible to interpret this tensor structure as a pure state-space geometry for the weighted Hilbert space with a modified inner product; however, we will not pursue such an interpretation in the present context.}. The LA-Berry curvature can be written down explicitly and is of the form 
\begin{equation}
\omega^{\text{LA}}_{ij}=\omega_{ij}+\frac{1}{L}\braket{\partial_{\{i}{\Psi(\theta)}|\mathbf{A}|\partial_{j\}}\Psi(\theta)}-\frac{1}{L^2}\partial_{\{i}L(\theta)\braket{{\Psi(\theta)}|\mathbf{A}|\partial_{j\}}\Psi(\theta)},
\label{LA-Berrycurvature}
\end{equation}
where we have assumed that the operator $\mathbf{A}$ does not explicitly depend on the (set of) parameters and used the notation $\{ij\}$ to denote the anti-symmetric combination of the given pair of indices. The LA-Berry curvature can also be considered as the corresponding curvature of the operator-weighted gauge connection of the form $\frac{1}{L}\braket{{\Psi(\theta)}|\mathbf{A}|\partial_{i}\Psi(\theta)}$.

\subsection{Position-space representation}
To understand the implications of the LA-QGT, in particular, in the context of classical information geometry, it is instructive to perform a position-space representation of the associated wavefunction and use the subsequent angular representation of it \cite{Facchi2010classical}. To this end, we obtain the Hermitian tensor
\begin{multline}
%\begin{split}
   {FS}^{\text{LA}}_{ij}=FS_{ij}+ \frac{1}{L}\int dx_{1}dx_{2}\sqrt{P_{1}P_{2}}e^{i\Phi_{21}}\tilde{A}_{12}\Big(\frac{1}{4}\partial_{i}\ln{P_{1}}\partial_{j}\ln{P_{2}}+\frac{i}{2}\partial_{i}\ln{P_{1}}\partial_{j}\Phi_{2}-\frac{i}{2}\partial_{i}\Phi_{1}\partial_{j}\ln{P_{2}}+\partial_{i}\Phi_{1}\partial_{j}\Phi_{2}\Big)\\
  - \frac{1}{L^2}\int dx_{1}dx_{2}dx_{3}dx_{4}\sqrt{P_{1}P_{2}P_{3}P_{4}}e^{i\Phi_{21}}e^{i\Phi_{43}}\tilde{A}_{12}\tilde{A}_{43}\Big(\frac{1}{4}\partial_{i}\ln{P_{1}}\partial_{j}\ln{P_{4}}+\frac{i}{2}\partial_{i}\ln{P_{1}}\partial_{j}\Phi_{4}-\frac{i}{2}\partial_{i}\Phi_{1}\partial_{j}\ln{P_{4}}+\partial_{i}\Phi_{1}\partial_{j}\Phi_{4}\Big),
%\end{split}
\end{multline}
where we have followed the notations of section \ref{classicalpullback}.
This is illuminating in the sense that the contribution of the `classical' part; the probability amplitude of the wavefunction $P(x;\theta)$ can be  thought to be a (non-local) deformation of the (position-space) FS tensor by the Kernel of the operator $\mathbf{A}$, where the standard FS tensor is obtained in the special case of $\tilde{A}(x_{a}, x_{b})=\delta(x_{a}-x_{b})$. At this point, we note that even though the real and symmetric part of the tensor $ {FS}^{\text{LA}}_{ij}$ defines the notion of distance and angle on the parameter manifold, the notion of a connection on this manifold, which defines  parallel-transport,  an essential concept for a generic curved manifold is only consistent with the trivial  metric connection \cite{Pal:2025anu}. To elaborate, the nature of the constraint imposed by the normalisation of the complex-valued wavefunction provides only the metric-compatible connection \cite{Heteneyi, Chen:2025lhm} on such geometries; thus, the powerful tool of duality of the $\pm\alpha$ connections used in the information geometry is not available.  To formulate such analogous $\pm\alpha$ connections in quantum systems with a non-trivial phase contribution in the wavefunction, we need to go beyond the standard Hermitian conjugation of the inner-product; a task we will consider in the next section.

%%%%%%%%%%%%%%%%%%%%%%%
\section{Biorthogonal loss-aware geometry}\label{secbiorthogonal}
\subsection{Gauge-invariant tensor structure}
In this section we will employ a biorthogonal inner-product to formulate the loss-aware geometry of the variational manifold, where the structures like the dual-connections of the standard information geometry can be induced. The overlap function to consider here is of the form \cite{Pal:2025anu}
\begin{equation}
\mathcal{D}^{\alpha}_{LA}(l_{1(\alpha)},l_{2(-\alpha)}) =\braket{l_{1(\alpha)}(\theta+\delta\theta)-l_{1(\alpha)}(\theta)|(\mathbf{I}+\frac{\mathbf{A}}{L_{\alpha}(\theta)})|l_{2(-\alpha)}(\theta+\delta\theta)-l_{2(-\alpha)}(\theta)},
\label{lossawraeoverlap}
\end{equation} 
where we have used the left state 
$ \bra{l_{1(\alpha)}(\theta)}$ and the right state, $ \ket{l_{2(-\alpha)}(\theta)}$, which have position space representations of the form 
\begin{equation}
\begin{split}
      l_{1(\alpha)}(x;\theta)=\frac{P^{\frac{1-\alpha}{2}}}{1-\alpha}e^{i(1-\alpha)\Phi}=\frac{\Psi^{1-\alpha}}{1-\alpha}~, ~~\text{and}~~\hspace{2mm}\\ l_{2(\alpha)}(x;\theta)=\frac{P^{\frac{1-\alpha}{2}}}{1-\alpha}e^{i(1+\alpha)\Phi}~, 
\end{split}
\end{equation}
for a real-valued parameter $\alpha\neq 1$.
They can be thought to be biorthogonal Hermitian conjugates of each other and are normalised in the sense of
\begin{equation}
    \braket{l_{1(\alpha)}|l_{2(-\alpha)}}=\braket{l_{2(-\alpha)}|l_{1(\alpha)}}=\frac{1}{1-\alpha^2}~,
\end{equation}
which can be set to unity after a trivial redefinition. This biorthogonal construction is  essentially a modification of the canonical inner-product on the Hilbert space, and throughout the rest of the paper, we have assumed that the operator we consider from now on is biorthogonal Hermitian $\mathbf{A}^{\#}=\mathbf{A}$. 
This condition, when written in terms of a set of left-right biorthogonal pair of complete basis (say, for example, the left-right eigenstates of a Hamiltonian, which is non-Hermitian with respect to the canonical inner-product on the Hilbert space) as $A_{nm}=A^{*}_{mn}$, in terms of the matrix elements in that basis \cite{Brodybiorthogonal}. Also, it should be noted that here, the cost function $L(\theta)$, is the expectation value of the operator with respect  to the biorthogonal pairing of states; $L_{\alpha}(\theta)= \braket{l_{1(\alpha)}(\theta)|\mathbf{A}|l_{2(-\alpha)}(\theta)}$. In the most generic case, for a non-Hermitian operator (for canonical inner product), the expectation value can, of course is not guaranteed to be real valued only. However, if we consider the so-called `associated' left-right states, then it is always real-valued, provided that the matrix elements of the operator satisfies the above criteria, as was considered in \cite{Brodybiorthogonal}, a condition that we will assume to be valid here also.

Expanding the overlap function \eqref{lossawraeoverlap} up to the quadratic order, we will recover the ($\alpha$)-FS geometry from the ``response'' of the identity operator for a small change of the parameter-space coordinates as 
\begin{equation}
\tilde{FS}^{(\alpha)}_{ij}=\braket{\partial_{i}{l_{1(\alpha)}}|\partial_{j}l_{2(-\alpha)}},
\label{alphaFStensorNI}
\end{equation}  

and as a ``response'' of the operator $\mathbf{A}$, we will obtain 
\begin{equation}
\tilde{FS}^{\text{LA}(\alpha)}_{ij}=\braket{\partial_{i}{l_{1(\alpha)}}|\frac{\mathbf{A}}{L_{\alpha}(\theta)})|\partial_{j}l_{2(-\alpha)}}~.
\label{LAalphaFStensorNI}
\end{equation}  
Even though these tensor structures transforms like  rank-$2$ tensors, under a change of coordinates $\{\theta_{i}\}$'s,  these are not physically consistent tensors, since they are not invariant under a phase-transformation of the associated states $ \bra{l_{1(\alpha)}(\theta)}$ and $ \ket{l_{2(-\alpha)}(\theta)}$, such that the norm remains invariant. However, using the standard procedure to fix this gauge-dependency issue \cite{Provost1980riemannian}, we obtain, 
\begin{equation}
FS^{(\alpha)}_{ij}=\braket{\partial_{i}{l_{1(\alpha)}}|\partial_{j}l_{2(-\alpha)}}-(1-\alpha^2)\braket{\partial_{i}l_{1(\alpha)}|l_{2(-\alpha)}}\braket{l_{1(\alpha)}|\partial_{j}l_{2(-\alpha)}}~,
\label{alphaFStensor}
\end{equation}  
as a tensor structure on the base manifold of the parameter space, where the set of parametrised states $\bra{l_{1(\alpha)}(\theta)}$ and $ \ket{l_{2(-\alpha)}(\theta)}$ are pulled-back to. On the other hand, similarly, we obtain the $U(1)$ gauge-invariant loss-aware tensor on the variational manifold, which is of the form
\begin{equation}
FS^{\text{LA}(\alpha)}_{ij}=\frac{1}{L_{\alpha}}\braket{\partial_{i}{l_{1(\alpha)}}|\mathbf{A}|\partial_{j}l_{2(-\alpha)}}-\frac{1}{L^{2}_{\alpha}(\theta)}\braket{\partial_{i}l_{1(\alpha)}|\mathbf{A}|l_{2(-\alpha)}}\braket{l_{1(\alpha)}|\mathbf{A}|\partial_{j}l_{2(-\alpha)}}~,
\label{LAalphaFStensor}
\end{equation}  
which we will refer to as the loss-aware $\alpha$-FS (LA-AFS) tensor structure in the subsequent sections. The position-space representation of this tensor is of the form
\begin{multline}
%\begin{split}
    FS^{\text{LA}(\alpha)}_{ij} =\\
    \frac{1}{L_{\alpha}} \Bigg( \int dx_{1}dx_{2}P_{1}^{\frac{1-\alpha}{2}}P_{2}^{\frac{1+\alpha}{2}}e^{i(1-\alpha)\Phi_{21}}\tilde{A}_{12}\Big(\frac{1}{4}\partial_{i}\ln{P_{1}}\partial_{j}\ln{P_{2}}+
    \frac{i(1-\alpha)}{2(1+\alpha)}\partial_{i}\ln{P_{1}}\partial_{j}\Phi_{2}-\frac{i}{2}\partial_{i}\Phi_{1}\partial_{j}\ln{P_{2}}+
\frac{(1-\alpha)}{(1+\alpha)}\partial_{i}\Phi_{1}\partial_{j}\Phi_{2}\Big)-\\
    \frac{1}{L_{\alpha}} \int dx_{1}dx_{2}dx_{3}dx_{4} (P_{1}P_{3})^{\frac{1-\alpha}{2}}(P_{2}P_{4})^{\frac{1+\alpha}{2}}\tilde{A}_{12}\tilde{A}_{34}e^{i(1-\alpha)\Phi_{21}}e^{i(1-\alpha)\Phi_{43}}\\
\Big(\frac{1}{4}\partial_{i}\ln{P_{1}}\partial_{j}\ln{P_{4}}+
    \frac{i(1-\alpha)}{2(1+\alpha)}\partial_{i}\ln{P_{1}}\partial_{j}\Phi_{4}-\frac{i}{2}\partial_{i}\Phi_{1}\partial_{j}\ln{P_{4}}+
\frac{(1-\alpha)}{(1+\alpha)}\partial_{i}\Phi_{1}\partial_{j}\Phi_{4}\Big)\Bigg)~.
\label{LAalphaFStensorpsoition}
%\end{split}
\end{multline}

To understand the role of this tensor \eqref{LAalphaFStensor}, we first take the ``classical'' limit; where we assume the (position-space) phase term of the wavefunction is trivial, as well as the operator kernel $\tilde{A}(x_{a}, x_{b})$ is real-valued. To this end, we obtain the following
\begin{multline}
%\begin{split}
    FS^{\text{LA-C}(\alpha)}_{ij} =\\
    \frac{1}{4L_{\alpha0}} \Bigg( \int dx_{1}dx_{2}P_{1}^{\frac{1-\alpha}{2}}P_{2}^{\frac{1+\alpha}{2}}\tilde{A}_{12}\partial_{i}\ln{P_{1}}\partial_{j}\ln{P_{2}}-
    \frac{1}{L_{\alpha0}} \int dx_{1}dx_{2}dx_{3}dx_{4} (P_{1}P_{3})^{\frac{1-\alpha}{2}}(P_{2}P_{4})^{\frac{1+\alpha}{2}}\tilde{A}_{12}\tilde{A}_{34}
\partial_{i}\ln{P_{1}}\partial_{j}\ln{P_{4}}\Bigg)~,
%\end{split}
\end{multline}
which represents a kernel-weighted non-local geometry on the space of PDFs associated with the position-space wave function.

\subsection{Decomposition of the non-Hermitian LA-AFS tensor}
The form \eqref{LAalphaFStensor} represents  a generic form of the non-Hermitian tensor structure (for the canonical inner-product) induced on the parameter manifold, based on the response of the test operator $\mathbf{A}$,  which can be decomposed into four individual tensors: \textbf{(a)} real \textit{and} symmetric part, \textbf{(b)} purely imaginary \textit{and} antisymmetric part, \textbf{(c)} real \textit{but} antisymmetric, \textbf{(d)} purely imaginary \textit{but} symmetric part.  

\paragraph{Real \textit{and} symmetric part:} From the LA-AFS tensor \eqref{LAalphaFStensor}, we can obtain the real as well as the symmetric rank-$2$ tensor, which has the form 
\begin{multline}
%\begin{split}
 g^{\text{LA}(\alpha)}_{ij} =\\
 \frac{1}{4L_{\alpha}} \Bigg( \int dx_{1}dx_{2}P_{1}^{\frac{1-\alpha}{2}}P_{2}^{\frac{1+\alpha}{2}} \Big(\cos{\Delta_{21}(\tilde{A}^{R}_{12}\tilde{\mathcal{B}}^{R}_{[ij](12)}-\tilde{A}^{I}_{12}\tilde{\mathcal{B}}^{I}_{[ij](12)})}-\sin{\Delta_{21}(\tilde{A}^{R}_{12}\tilde{\mathcal{B}}^{I}_{[ij](12)}+\tilde{A}^{I}_{12}\tilde{\mathcal{B}}^{R}_{[ij](12)})}\Big) \\
-\frac{1}{L_{\alpha}} \int dx_{1}dx_{2}dx_{3}dx_{4}(P_{1}P_{3})^{\frac{1-\alpha}{2}}(P_{2}P_{4})^{\frac{1+\alpha}{2}} \Big(\cos{\Theta}(\tilde{C}^{R}_{12,34}\tilde{\mathcal{B}}^{R}_{[ij](12)}-\tilde{C}^{I}_{12,34}\tilde{\mathcal{B}}^{I}_{[ij](12)})-\sin{\Theta}(\tilde{C}^{R}_{12,34}\tilde{\mathcal{B}}^{I}_{[ij](12)}+\tilde{C}^{I}_{12,34}\tilde{\mathcal{B}}^{R}_{[ij](12)})\Big)
\Bigg),
\label{alphametric}
%\end{split}
\end{multline}
where we have used the notation $\Delta_{ba}=(1-\alpha)\Phi_{ba}$, $\tilde{A}_{ab}=\tilde{A}^{R}_{ab}+i\tilde{A}^{I}_{ab}$, $\tilde{\mathcal{B}}^{R}_{ij(ab)}=\Big(\frac{1}{4}\partial_{i}\ln{P_{a}}\partial_{j}\ln{P_{b}}+\frac{(1-\alpha)}{(1+\alpha)}\partial_{i}\Phi_{a}\partial_{j}\Phi_{b}\Big)$, $\tilde{\mathcal{B}}^{I}_{ij(ab)}=\Big(\frac{(1-\alpha)}{2(1+\alpha)}\partial_{i}\ln{P_{a}}\partial_{j}\Phi_{b}-\frac{1}{2}\partial_{i}\Phi_{a}\partial_{j}\ln{P_{b}}\Big)$, $\Theta=\Delta_{21}+\Delta_{43}$, $\tilde{C}^{R}_{12,34}+i\tilde{C}^{I}_{12,34}=\tilde{A}_{12}\tilde{A}_{34}$  and $[ij]$ denotes the symmetrisation with respect to the pair of indices. As can be seen, in the specific case of the identity operator, for the delta function kernel, \eqref{alphametric} reduces to the following 
\begin{equation}
    g^{(\alpha)}_{ij}=\frac{1}{4}\mathcal{E}_{p}\Big[\partial_{i}\ln{P}\partial_{j}\ln{P}\Big]+\frac{(1-\alpha)}{(1+\alpha)}\Bigg(\mathcal{E}_{p}\Big[\partial_{i}\Phi\partial_{j}\Phi\Big]-\mathcal{E}_{p}\Big[\partial_{i}\Phi\Big]\mathcal{E}_{p}\Big[\partial_{j}\Phi\Big]\Bigg)~,
\end{equation}
which is the standard contribution form the real and symmetric part to the QMT in the biorthogonal setting we are using \cite{Pal:2025anu}.

\paragraph{Purely imaginary \textit{and} antisymmetric part:} 
After proper asymmetrisation and taking the imaginary part, we have,
\begin{multline}
%\begin{split}
    \omega^{\text{LA}(\alpha)}_{ij} =\\
    \frac{1}{4L_{\alpha}} \Bigg( \int dx_{1}dx_{2}P_{1}^{\frac{1-\alpha}{2}}P_{2}^{\frac{1+\alpha}{2}} \Big(\sin{\Delta_{21}}(\tilde{A}^{R}_{12}\tilde{\mathcal{B}}^{R}_{\{ij\}(12)}-\tilde{A}^{I}_{12}\tilde{\mathcal{B}}^{I}_{\{ij\}(12)})+\cos{\Delta_{21}}(\tilde{A}^{R}_{12}\tilde{\mathcal{B}}^{I}_{\{ij\}(12)}+\tilde{A}^{I}_{12}\tilde{\mathcal{B}}^{R}_{\{ij\}(12)})
    \Big)-\\
   \frac{1}{L_{\alpha}} \int dx_{1}dx_{2}dx_{3}dx_{4}(P_{1}P_{3})^{\frac{1-\alpha}{2}}(P_{2}P_{4})^{\frac{1+\alpha}{2}} \Big(\sin{\Theta}(\tilde{C}^{R}_{12,34}\tilde{\mathcal{B}}^{R}_{\{ij\}(12)}-\tilde{C}^{I}_{12,34}\tilde{\mathcal{B}}^{I}_{\{ij\}(12)})+\cos{\Theta}(\tilde{C}^{R}_{12,34}\tilde{\mathcal{B}}^{I}_{\{ij\}(12)}+\tilde{C}^{I}_{12,34}\tilde{\mathcal{B}}^{R}_{\{ij\}(12)})\Big) \Bigg).
%\end{split}
\end{multline}
Similarly as above we get for the identity operator, the form of this tensor
\begin{equation}
\omega^{(\alpha)}_{ij}=\frac{i}{2(\alpha+1)}\mathcal{E}_{p}\Big[\partial_{i}\ln{P}~\partial_{j}\Phi-\partial_{i}\Phi~\partial_{j}\ln{P} \Big],
\label{berrycuravturealpha}
\end{equation}
which is the  contribution of the purely imaginary and anti-symmetric term in the full Berry curvature.

\paragraph{Real \textit{but} antisymmetric part:} 
Due to the particular form of the tensor used in this analysis, the overall contribution to the variational geometry in this case will contain two additional contributions; first one is the real but the antisymmetric part of the LA-AFS tensor and is of the form,
\begin{multline}
%\begin{split}
 \bar{\omega}^{\text{LA}(\alpha)}_{ij} =\\
 \frac{1}{4L_{\alpha}} \Bigg( \int dx_{1}dx_{2}P_{1}^{\frac{1-\alpha}{2}}P_{2}^{\frac{1+\alpha}{2}} \Big(\cos{\Delta_{21}(\tilde{A}^{R}_{12}\tilde{\mathcal{B}}^{R}_{\{ij\}(12)}-\tilde{A}^{I}_{12}\tilde{\mathcal{B}}^{I}_{\{ij\}(12)})}-\sin{\Delta_{21}(\tilde{A}^{R}_{12}\tilde{\mathcal{B}}^{I}_{\{ij\}(12)}+\tilde{A}^{I}_{12}\tilde{\mathcal{B}}^{R}_{\{ij\}(12)})}\Big) \\
-\frac{1}{L_{\alpha}} \int dx_{1}dx_{2}dx_{3}dx_{4}(P_{1}P_{3})^{\frac{1-\alpha}{2}}(P_{2}P_{4})^{\frac{1+\alpha}{2}} \Big(\cos{\Theta}(\tilde{C}^{R}_{12,34}\tilde{\mathcal{B}}^{R}_{\{ij\}(12)}-\tilde{C}^{I}_{12,34}\tilde{\mathcal{B}}^{I}_{\{ij\}(12)})-\sin{\Theta}(\tilde{C}^{R}_{12,34}\tilde{\mathcal{B}}^{I}_{\{ij\}(12)}+\tilde{C}^{I}_{12,34}\tilde{\mathcal{B}}^{R}_{\{ij\}(12)})\Big)
\Bigg).
%\end{split}
\end{multline}
It is to be noted that, for the for the response of the identity operator, the contribution of this term for the LA-AFS geometry identically vanishes.

\paragraph{Purely imaginary \textit{but} symmetric part:} 
Another novel implication of the use of the structure we are using, is the presence of the imaginary but symmetric term;
\begin{multline}
%\begin{split}
    \bar{g}^{\text{LA}(\alpha)}_{ij} =\\
    \frac{1}{4L_{\alpha}} \Bigg( \int dx_{1}dx_{2}P_{1}^{\frac{1-\alpha}{2}}P_{2}^{\frac{1+\alpha}{2}} \Big(\sin{\Delta_{21}}(\tilde{A}^{R}_{12}\tilde{\mathcal{B}}^{R}_{[ij](12)}-\tilde{A}^{I}_{12}\tilde{\mathcal{B}}^{I}_{[ij](12)})+\cos{\Delta_{21}}(\tilde{A}^{R}_{12}\tilde{\mathcal{B}}^{I}_{[ij](12)}+\tilde{A}^{I}_{12}\tilde{\mathcal{B}}^{R}_{\{ij\}(12)})
    \Big)-\\
   \frac{1}{L_{\alpha}} \int dx_{1}dx_{2}dx_{3}dx_{4}(P_{1}P_{3})^{\frac{1-\alpha}{2}}(P_{2}P_{4})^{\frac{1+\alpha}{2}} \Big(\sin{\Theta}(\tilde{C}^{R}_{12,34}\tilde{\mathcal{B}}^{R}_{[ij](12)}-\tilde{C}^{I}_{12,34}\tilde{\mathcal{B}}^{I}_{[ij](12)})+\cos{\Theta}(\tilde{C}^{R}_{12,34}\tilde{\mathcal{B}}^{I}_{[ij](12)}+\tilde{C}^{I}_{12,34}\tilde{\mathcal{B}}^{R}_{[ij](12)})\Big) \Bigg).
%\end{split}
\end{multline}
Importantly, however, in the limit of the delta-function kernel this reduces to 
\begin{equation}
    \tilde{g}^{(\alpha)}_{ij}=-\frac{i\alpha}{2(1+\alpha)}\mathcal{E}_{p}\Big[\partial_{i}\ln{P}~\partial_{j}\Phi+\partial_{i}\Phi~\partial_{j}\ln{P}\Big]~,
\label{flippedsymmetric}
\end{equation}
indicating the non-trivial amplitude-phase mixing in the $\alpha\neq 0$ case.

%%%%%%%%%
\subsection{Loss-aware $\pm\alpha$-connections}
One of the advantages of the biorthogonal formalism we have adapted here is that, it is possible to define a notion of the $\pm\alpha$-connections on the variational manifold; in a similar way as that of the classical information geometry, after properly removing the gauge ambiguities. Collecting the third-order terms in the expansion of the loss-aware overlap \eqref{lossawraeoverlap}, we obtain two rank-$3$ (non-)tensors, which are of the form 
\begin{equation}
\tilde{\Gamma}_{ij,k}^{\text{LA}-1(\alpha)}=\braket{\partial_{i}\partial_{j}{l_{1(\alpha)}}|\frac{\mathbf{A}}{L_{\alpha}}|\partial_{k}l_{2(-\alpha)}}, 
\label{LAalphaconnection}
\end{equation}
\text{and}
  \begin{equation}
\tilde{\Gamma}_{ij,k}^{\text{LA}-2(-\alpha)}=\braket{\partial_{k}{l_{1(\alpha)}}|\frac{\mathbf{A}}{L_{\alpha}}|\partial_{i}\partial_{j}l_{2(-\alpha)}}~,
\label{LAalphamiconnection}
\end{equation} 
and is not physically meaningful due to the inherent gauge dependencies. The proper physically meaningful form of these two objects, which can be easily verified to be not tensors; but are physically relevant connection coefficients on the manifold and is symmetric in two indices, to be of the form 
%\begin{align}
\begin{multline}
%\begin{split}
   \Gamma_{ij,k}^{\text{LA}-1(\alpha)}=\frac{1}{L_{\alpha}}\braket{\partial_{i}\partial_{j}{l_{1(\alpha)}}|\mathbf{A}|\partial_{k}l_{2(-\alpha)}}-\frac{1}{L^{2}_{\alpha}}\braket{\partial_{i}\partial_{j}l_{1(\alpha)}|\mathbf{A}|l_{2(-\alpha)}}\braket{l_{1(\alpha)}|\mathbf{A}|\partial_{k}l_{2(-\alpha)}}-\frac{2}{L^{2}_{\alpha}}\braket{\partial_{[i}l_{1(\alpha)}|\mathbf{A}|l_{2(-\alpha)}}\braket{\partial_{j]}l_{1(\alpha)}|\mathbf{A}|\partial_{k}l_{2(-\alpha)}}\\
   +\frac{2}{L^{3}_{\alpha}}\braket{\partial_{[i}l_{1(\alpha)}|\mathbf{A}|l_{2(-\alpha)}}\braket{\partial_{j]}l_{1(\alpha)}|\mathbf{A}|l_{2(-\alpha)}}\braket{l_{1(\alpha)}|\mathbf{A}|\partial_{k}l_{2(-\alpha)}}~,
\label{LAalphaconnectionGI} 
%\end{split}
\end{multline} 
%\end{align}
and 
\begin{multline}
%\begin{split}
   \Gamma_{ij,k}^{\text{LA}-2(-\alpha)}=\frac{1}{L_{\alpha}}\braket{\partial_{k}{l_{1(\alpha)}}|\mathbf{A}|\partial_{i}\partial_{j}l_{2(-\alpha)}}-\frac{1}{L^{2}_{\alpha}}\braket{\partial_{k}l_{1(\alpha)}|\mathbf{A}|l_{2(-\alpha)}}\braket{l_{1(\alpha)}|\mathbf{A}|\partial_{i}\partial_{j}l_{2(-\alpha)}}-\frac{2}{L^{2}_{\alpha}}\braket{l_{1(\alpha)}|\mathbf{A}|\partial_{[i}l_{2(-\alpha)}}\braket{\partial_{k}l_{1(\alpha)}|\mathbf{A}|\partial_{j]}l_{2(-\alpha)}}\\
   +\frac{2}{L^{3}_{\alpha}}\braket{\partial_{k}l_{1(\alpha)}|\mathbf{A}|l_{2(-\alpha)}}\braket{l_{1(\alpha)}|\mathbf{A}|\partial_{[i}l_{2(-\alpha)}}\braket{l_{1(\alpha)}|\mathbf{A}|\partial_{j]}l_{2(-\alpha)}}~.
\label{LAalphaconnectionmGI} 
%\end{split}
\end{multline}
From the expression \eqref{LAalphaconnectionGI}, it can be easily checked that in the special case when the phase of the wavefunction in question is trivial, we have 
\begin{multline}
%\begin{split}
\Gamma_{ij,k}^{\text{LA}-1(\alpha)}|_{C}=\frac{1}{4}\int dx_{1}dx_{2}P_{1}^{\frac{1-\alpha}{2}}P_{2}^{\frac{1+\alpha}{2}} \tilde{A}_{12}~\Bigg(\partial_{i}\partial_{j}\ln{P_{1}}+\frac{1-\alpha}{2}\partial_{i}\ln{P_{1}}\partial_{j}\ln{P_{1}}~\Bigg)\partial_{k}\ln{P_{2}}-\\
\frac{1}{(1-\alpha^2)L_{\alpha 0}^2}\int dx_{1}dx_{2}P_{1}^{\frac{1-\alpha}{2}}P_{2}^{\frac{1+\alpha}{2}}\tilde{A}_{12}\partial_{k}\ln{P_{2}}\int dx_{3}dx_{4}P_{3}^{\frac{1-\alpha}{2}}P_{4}^{\frac{1+\alpha}{2}} \tilde{A}_{34}~\Bigg(\partial_{i}\partial_{j}\ln{P_{3}}+\frac{1-\alpha}{2}\partial_{i}\ln{P_{3}}\partial_{j}\ln{P_{3}}\Bigg)-\\
\frac{1}{(1+\alpha)L_{\alpha 0}^2}\int dx_{1}dx_{2}P_{1}^{\frac{1-\alpha}{2}}P_{2}^{\frac{1+\alpha}{2}}\tilde{A}_{12}\partial_{[i}\ln{P_{1}}\int dx_{3}dx_{4}P_{3}^{\frac{1-\alpha}{2}}P_{4}^{\frac{1+\alpha}{2}}\tilde{A}_{34}\partial_{j]}\ln{P_{3}}\partial_{k}\ln{P_{4}}+\\
\frac{1}{(1+\alpha)(1-\alpha^2)L_{\alpha 0}^3}\int dx_{1}dx_{2}P_{1}^{\frac{1-\alpha}{2}}P_{2}^{\frac{1+\alpha}{2}}\tilde{A}_{12}\partial_{[i}\ln{P_{1}}\int dx_{3}dx_{4}P_{3}^{\frac{1-\alpha}{2}}P_{4}^{\frac{1+\alpha}{2}}\tilde{A}_{34}\partial_{j]}\ln{P_{4}}\int dx_{5}dx_{6}P_{5}^{\frac{1-\alpha}{2}}P_{6}^{\frac{1+\alpha}{2}}\tilde{A}_{56}\partial_{k}\ln{P_{6}}. 
\label{eqgamma1zerophase}
%\end{split}
\end{multline}
Furthermore, in the specific case, when the kernel represents the delta function kernel of the identity operator, from \eqref{eqgamma1zerophase} we get back,
\begin{equation}
    \Gamma^{(\alpha)}_{ij,k}(\theta)= \frac{1}{4}\int dx P ~\Bigg(\partial_{i}\partial_{j}\ln{P}+\frac{1-\alpha}{2}\partial_{i}\ln{P}~\partial_{j}\ln{P}\Bigg)~\partial_{k}\ln{P},
\end{equation}
which is precisely the $+\alpha$ connection of the classical information geometry \cite{Amari2000methods, Amari2016information, Nagaoka2026differential}, with the factor $\frac{1}{4}$ coming from the definition of the quantum fisher information metric. Similarly, expanding the corresponding expression \eqref{LAalphaconnectionmGI} and considering  the special classical as well as the delta-function kernel case, we will have, 
\begin{equation}
    \Gamma^{(-\alpha)}_{ij,k}(\theta)= \frac{1}{4}\int dx P ~\Bigg(\partial_{i}\partial_{j}\ln{P}+\frac{1+\alpha}{2}\partial_{i}\ln{P}~\partial_{j}\ln{P}\Bigg)~\partial_{k}\ln{P},
\end{equation}
which is the standard form of the $-\alpha$ connection. The preceding discussion thus places the LA-geometry considered in section \ref{LA-QMT-1} into a firmer ground; since we can now have a first-principle construction of various types of loss-dependent geometries as  the systematic expansion of the response of concerned operators, by considering proper overlap integrals.

%%%%%%%%%%%%%%%%
\section{Conclusions and discussion}
The efficiency of the natural gradient descent for cost optimisation in classical and quantum systems has its roots in the geometry-aware formulation of the same, where the Riemannian Fisher information metric determines the direction of the natural gradient flow on such a curved manifold. Even though the optimisers using the natural gradient update rule have in-built information about the geometry of the space of the probability distribution under consideration, they are not sensitive to the underlying geometry of the space of possible outcomes, which for quantum systems can be roughly thought to be the outcome of measurement of an observable with respect to a quantum state on the Hilbert space. Since the quantum variational eigensolver, which is the state-of-the-art algorithm for the currently available near-term quantum computers, uses a classical optimiser having support on a classical space of possible outcomes to update the circuit parameter for the next iteration in order to optimise the cost function, it is expected that the knowledge of the geometry of the same would be beneficial for extracting possible computational advantages. 

In this work, we have proposed such a scheme for a natural gradient-type optimiser, but instead of the Fisher information metric of the space of classical probability distributions or the quantum metric tensor for parametrised quantum states, we have used a novel pull-back metric on the parameter manifold, which is induced on the loss landscape when it is embedded in an ambient space, which in turn is a product manifold of the standard classical or the quantum statistical manifold, augmented by an Euclidean direction, governed by the functional dependence of the loss function. We have systematically constructed such a pull-back-metric geometry both for classical and quantum cases, which in the latter case imposes a two-point amplitude-phase mixing structure through the non-local operator kernel. The resulting natural gradient descent flow controlled by the pull-back metric is effectively a neat and conceptually pleasing form of the gradient clipping method, which regularises the gradient vector in the higher curvature regime \cite{Harvey2025optimiser}. 

In order to increase the efficiency of the loss-aware gradient descent in the flatter region without significantly compromising with the stability in the higher-curvature regions of the loss landscape, next we consider a scaled version of the loss-aware geometry, where not only the geometry is anisotropically stretched or shrinks, but also is zoomed in or out by means of a conformal transformation. We considered three specific examples of this conformal class of loss-ware geometry controlled by a single parameter $\gamma$, and discussed how  effective learning rates can be controlled by changing the conformal factor. 
To glean the practical advantages or disadvantages of these modifications of the geometry, we performed the gradient-based energy minimisation schemes to minimise the energy of a random $4$-dimensional Hamiltonian acting on $2$ qubits, both with or without the presence of external noise. We adapted a block-diagonal approximation of the full quantum geometric tensor and performed the different cases of optimisers discussed in this work. In the presence of  perturbative noise in the block-diagonal components of the quantum geometric tensor, we observe that QNG performs most effectively during the early stages of optimisation and provides the most robust convergence overall. However, loss-adaptive conformally modified variants, particularly CLA-$3$-QNG, more frequently achieve the best convergence times and exhibit superior performance in favourable instances. In the noise-free regime, on the other hand, the median performance of all methods is broadly comparable; however, the conformally adapted schemes demonstrate slightly improved best-case performance as compared to the noisy case, achieving higher win rates and more frequently attaining the optimal convergence time compared to standard QNG. 
For the classical optimisation scenario, curvature-aware methods that incorporate K-FAC preconditioning show a clear advantage over Adam and SGD-RMS, achieving lower training loss and reaching target validation accuracy in substantially fewer epochs. The baseline optimisers converge more slowly and exhibit similar behaviour, reflecting their limited adaptivity compared to methods that account for the underlying geometry of the loss landscape. Among all evaluated approaches, F-CLA-$3$ performs best, closely followed by F-LANG, underscoring the effectiveness of curvature-aware optimisation.

To have a concrete understanding of the role of these loss-adaptive versions of the parameter manifold geometries, in particular to investigate if it is possible to formulate a notion of dual connections in such situations, we constructed a novel version of these geometries starting from a well-defined overlap of states and the corresponding response of an operator in those perturbed states. We constructed the associated quantum metric tensor as well as the Berry curvature from the Hermitian tensor structure, which, however, only provides the notion of metric-compatible connections in such a manifold. We resorted to a biorthogonal decomposition of the initial (position-space) wavefunction and developed a consistent formulation of a non-Hermitian tensor structure and two connections from the second- and third-order properties of a biorthogonal overlap function, respectively. Our results thus place the construction of such  loss-aware geometries into an information geometric context, which we hope will be useful in further explorations to this end.

%%%%%%
\begin{center}
	\bf{Acknowledgments}
\end{center}
We would like to thank Kuntal Pal for  discussion and comments on the draft.  The work of Kunal Pal is  supported by the YST Programme at the APCTP through the Science and
Technology Promotion Fund and the Lottery Fund of the Korean Government. This was also supported by the Korean Local Governments -
Gyeongsangbuk-do Province and Pohang City. Ankit Gill is thankful for the financial support received from the FARE Fellowship at IIT Kanpur.

%%%%%%%%%%%%%%%%%%%%%
\appendix
\section{Comparison of different geometries for the exponential family of probability distribution}\label{NGcomparisonexponential}
In this appendix, we will provide a comparison of different geometries and the corresponding classical NG-schemes proposed in this paper for a simple choice of PDF, the exponential family, with a Gaussian kernel, which can be thought of as the kernel associated with the heat semigroup operator. In this simplified case, it is possible to obtain the matrix elements of the LA-metrics, the trajectories in the $\theta^{1}- \theta^{2}$ plane in a simple analytical form, as well as the implicit form of the trajectory  $\theta^{2}(t)$ in some cases. The primary aim of this analysis is to see if it is possible to obtain insights about the possible change of the metric and associated geometric quantities, such as scalar curvatures, in an analytically tractable way due to the presence of the effective rank-$1$ deformation of the natural geometry of the statistical manifold, either described by the FIM or  QMT. Specifically, our focus will be to check how the hyperbolic nature of the statistical manifold equipped with FIM is changed due to the deformation. Even though the GD equations being first-order equations does not get affected by the curvature of the manifold directly, the implications of a possible change of the geometry of the manifold are expected to significantly change the geometric quantities associated with the metric; see, for example, \cite{TS2, TS1, Gill:2025upp}. 
For the FIM of this family, it is well known that the manifold is a space of constant, negative curvature, a manifestation of the scale-invariant nature of the PDF. To be more precise, a PDF can be thought to have a symmetry if a corresponding change of the parameters specifying the PDF $(\{\theta\} \rightarrow \{\tilde{\theta}\})$ can be absorbed in a corresponding change of the sample-space variables $(\{x\} \rightarrow \{\tilde{x}\})$, such that $\int dx ~P(x, \tilde{\theta}) =\int d\tilde{x} ~P(\tilde{x}, \theta)$ \cite{Erdmenger2020information}. This symmetry of the PDF might enforce stronger requirements on the geometry; i.e., the symmetry of the PDFs is typically manifest in the corresponding information metric, and we might need other geometric structures on the statistical manifold to distinguish such metrics with the same symmetry groups induced from different PDFs \cite{Pal:2022szb}.
Then the natural question to ask is: if such a statement holds for the kernel-weighted geometries induced by the pull-back on the parameter manifold, say, for example, in the classical case, eq. \eqref{eq:classicalpullback}. To incorporate the necessary scaling structure of the observable-space parameters (henceforth collectively denoted as $\{\xi\}$), we will consider a joint symmetry transformation of $(\{\theta\} \rightarrow \{\tilde{\theta}\})$ and $(\{\xi\} \rightarrow \{\tilde{\xi}\})$ such that the tensor density associated with the integral measure is invariant:  
\begin{equation}
    \int~dx_{1}dx_{2}\sqrt{P(x_{1}, \tilde{\theta})P(x_{2}, \tilde{\theta})}~A(x_{1}, x_{2}, \tilde{\xi})=\int~d\tilde{x}_{1}d\tilde{x}_{2}\sqrt{P(\tilde{x}_{1}, \theta)P(\tilde{x}_{2},\theta)}~A(\tilde{x}_{1}, \tilde{x}_{2}, \xi).
\end{equation}
As we will see later, these scaling transformations precisely constrain the LA-metric to be governed by a single scaling parameter for the particular form of the loss function we will consider. 

Before presenting the explicit solutions in each case, let us first comment on some generic features of the solutions that we will discuss in the sequel. First of all, since the LA modifications or the corresponding conformal modifications do not change the direction of the gradient flow, only the effective step sizes, the descent trajectories  are straight lines through the origin, in all the five cases we have discussed in this paper; as we will explicitly see later. Similarly, due to the $\theta^{1}$-independent nature of the loss function, which can be traced back to the translation symmetry of the kernel used, the equation of motion (EOM) for $\theta^{2}(t)$ represents an autonomous system and, in principle, is integrable. Another important  characteristic of each class of metric we have considered is the associated curvature scalar, which in the simplified two-dimensional parameter space we are considering has only one independent component. 

To be more quantitative, let us first record the cost function for this choice of PDF, which is of the form \eqref{ExponentialPDF}, can be analytically computed in a closed form $L(\theta^{2})=\sqrt{\frac{2}{\Delta}}$, for $\Delta=2-\kappa^2\theta^{2} > 2$, where $\kappa$ is the width of the non-local Gaussian kernel chosen. The variational problem for this cost function is ill-defined, since there is no minimum of the same within the finite range of the trial parameters, indications of the spectral nature of the heat semigroup operator chosen. However, our motivation here for performing this exercise is to explicitly demonstrate how the loss-aware geometry affects the symmetry and the invariant quantities associated with the FIM, as we indicated above.

\textbf{Case-1: FIM}~~For the chosen form of the exponential family, FIM in the canonical coordinates $\theta^{1}, \theta^{2}$, we can write down the components of the FIM solely from the potential function in this case and it is of the form:
\begin{equation}
 g^{(\text{FIM})}_{ij}=
 \begin{bmatrix}
-\frac{1}{2\theta^{2}} & \frac{\theta^{1}}{2
(\theta^{2})^2} \\
\frac{\theta^{1}}{2
(\theta^{2})^2}  & \frac{1}{2
(\theta^{2})^2} -\frac{(\theta^{1})^{2}}{2
(\theta^{2})^3} 
\end{bmatrix}.
\label{FIMmatrixform}
\end{equation}
As can be easily checked, the Ricci curvature scalar for  this metric \eqref{FIMmatrixform} is $R_{\text{FIM}}=-1$. Another important feature of this geometry is that it describes a scale-invariant geometry, which can be best seen by considering the line element in the original $(\mu, \delta)$ coordinates, which is diagonal now and is of the form: $\text{d}s^2=\frac{\text{d}\mu^2+2\text{d}\delta^2}{\delta^2}$ and this remains the same even after a rescaling of the coordinates. Furthermore, the metric is also invariant under a translation of the mean $\mu \rightarrow \mu+\mu_{0}$; therefore, the isometry group of this geometry is generated by  scaling \textit{and} translation in $\mu$ and simultaneous scaling of $\delta$.

The gradient descent trajectories with respect to the FIM corresponding to the PDF (which, of course, does not depend on the cost function) can be written down as 
\begin{equation}
    \frac{d\theta^{1}}{dt}=-\eta g^{(\text{FIM})12}\partial_{2}L(\theta)=-\theta^{1}\theta^{2}\Omega~ \text{and} ~\frac{d\theta^{2}}{dt}=-\eta g^{(\text{FIM})22}\partial_{2}L(\theta)~=-(\theta^{2})^2\Omega,
\label{eqtheta12FIM}
\end{equation}
where we have denoted $\partial_{\theta^{2}}L(\theta)$ as $\partial_{2}L(\theta)$ and $\Omega(\theta^{2})=2\eta\sqrt{\frac{\pi\kappa^6}{\Delta^3}}$. From these EOMs, it is evident that in the $\theta^{1}- \theta^{2}$ plane, the trajectories are straight lines passing through the origin, where the slope is determined by the initial values of $\theta^{1}(t=0)$ and  $ \theta^{2}(t=0)$. Moreover, in this simplified case, we can integrate the autonomous equation for $\theta^{2}$ to obtain the implicit solution of the form 
\begin{equation}
    t= \frac{1}{ \eta\sqrt{2\pi}\kappa}\Big(\frac{(2\omega-1)}{\omega}\sqrt{1+\omega}-3\operatorname{arccoth}{(\sqrt{1+\omega})}\Big),
\end{equation}
with $\omega=-\frac{\kappa^2}{2}\theta^{2}$, which implicitly shows the update steps of $\Delta\theta^{2}$ for minimisation.

\textbf{Case-2: LA-metric}~~ For the induced metric on the parameter manifold, of the form \eqref{LA-metric}, for this family of distribution and the Gaussian kernel can be analytically computed to be of the form
\begin{equation}
 g^{(\text{LA})}_{ij}=
 \begin{bmatrix}
-\frac{1}{2\theta^{2}} & \frac{\theta^{1}}{2
(\theta^{2})^2} \\
\frac{\theta^{1}}{2
(\theta^{2})^2}  & \frac{1}{2
(\theta^{2})^2} -\frac{(\theta^{1})^{2}}{2
(\theta^{2})^3}+\Sigma(\theta^{2})
\end{bmatrix},
\label{LA-metricmatrixform}
\end{equation}
where $\Sigma(\theta^{2})=\frac{\kappa^4}{2\Delta^3}$. For this metric, we can compute the form of the Ricci scalar curvature $R_{\text{LA}}=\frac{2\Sigma (\theta^{2})^2+2\Sigma^{\prime} (\theta^{2})^3-1}{(1+2\Sigma(\theta^{2})^2)^{2}}$, where an over-prime denotes the partial derivative with respect to $\theta^{2}$. Importantly, the joint symmetries of the normal distribution and that of the Gaussian kernel used suggest that the resulting pull-back geometry is still scale-invariant, a direct manifestation of the underlying symmetries. This again can be understood from the form of the line element in the $(\mu, \delta)$ coordinates, which now has the form $\text{d}s_{\text{{LA}}}^2=\frac{\text{d}\mu^2+2\text{d}\delta^2}{\delta^2}+\frac{4\kappa^{4}}{(4\delta^2+\kappa^2)^{3}}\text{d}\delta^2$ and it is evident that the final term preserves the translation symmetry of $\mu$,  as well as the scaling symmetries of $(\mu, \delta)$, provided the width of the Gaussian kernel $\kappa$ is also scaled accordingly. It should be noted however, that the scale transformation associated with the observable-space parameter $\kappa$ is \textit{not}  a coordinate transformation on the statistical manifold in this context; rather, it should be understood as a simultaneous reparametrisation (-invariance) of the curve $g(\kappa)$ in the space of metrics for the LA-geometries. As a result of scaling invariance, the scalar curvature in this case $R_{\text{LA}}$ depends only on the effective emergent scale $\beta=\frac{\kappa}{\delta}$ and is of the form $R_{\text{LA}}=-\frac{(2+\frac{\beta^2}{2})^2}{(8+\frac{\beta^6}{8}+\frac{7\beta^4}{4}+6\beta^2)^2}\Big(16(1+\beta^2)+\frac{\beta^8}{16}+\frac{5\beta^4}{4}\Big)$. As is immediately clear, the curvature for the LA geometry is everywhere hyperbolic; thus, the rank-$1$ deformations of the FIM in this case preserve the hyperbolic nature of the same. As before, we can write down the gradient descent trajectories for the LA-metric, which are now of the form 
\begin{equation}
    \frac{d\theta^{i}_{\text{LA}}}{dt}=\frac{1}{(1+2(\theta^{2})^{2}\Sigma)}\frac{d\theta^{i}_{\text{FIM}}}{dt},
\label{theta12LA}
\end{equation}
written in terms of the trajectories for the FIM in \eqref{eqtheta12FIM} and again describes straight lines passing through the origin.

\textbf{Case-3: CLA metrics}~~
For the general family of CLA metrics of the form \eqref{CLAmetric}, the Ricci curvature scalar can be computed similarly in terms of the conformal factor $C(\theta^{2})$  and is of the form:
\begin{equation}
    R_{\text{CLA}}=\frac{e^{-C}}{\big(1+2(\theta^{2})^{2}\Sigma\big)^{2}}\Bigg(2(\theta^{2})^{2}\Sigma+2(\theta^{2})^{3}\Sigma^{\prime}-2(\theta^{2})^{2}C^{\prime\prime}\big(1+2(\theta^{2})^{2}\Sigma\big)-\theta^{2}C^{\prime}\Big(3+2(\theta^{2})^{2}\Sigma-2(\theta^{2})^{3}\Sigma^{\prime}\Big)-1\Bigg),
\end{equation}
with the conformal factor $C$ determining the explicit form of each type of geometries. To exemplify, for the choice of conformal factor  $C{(\theta^{2})}=\gamma\log\Big(1+g^{(\text{FIM})ij}\partial_{i}L(\theta^{2})\partial_{j}L(\theta^{2})\Big)$, we will have the CLA-1 type geometries considered in section \ref{CLA-1}, where the line-element is
$\text{d}s_{\text{CLA-1}}^2=\Big(1+\frac{\kappa^4\delta^2}{(4\delta^2+\kappa^2)^{3}}\Big)^{\gamma}\Big(\frac{\text{d}\mu^2+2\text{d}\delta^2}{\delta^2}+\frac{4\kappa^{4}}{(4\delta^2+\kappa^2)^{3}}\text{d}\delta^2\Big)$. The scale-invariance of the metric, induced by the scaling transformations of the joint probability and observable-space parameters, is again manifest in the metric written in the $\mu, \delta$ coordinates. We can obtain the form of the Ricci scalar curvatures for the other CLA-type metrics in a similar way for each type of conformal factors used; which we omit for brevity. 

\begin{figure}[h!]
	\begin{minipage}{0.4\linewidth}
		\centering
\includegraphics[width=2.8in,height=2.5in]{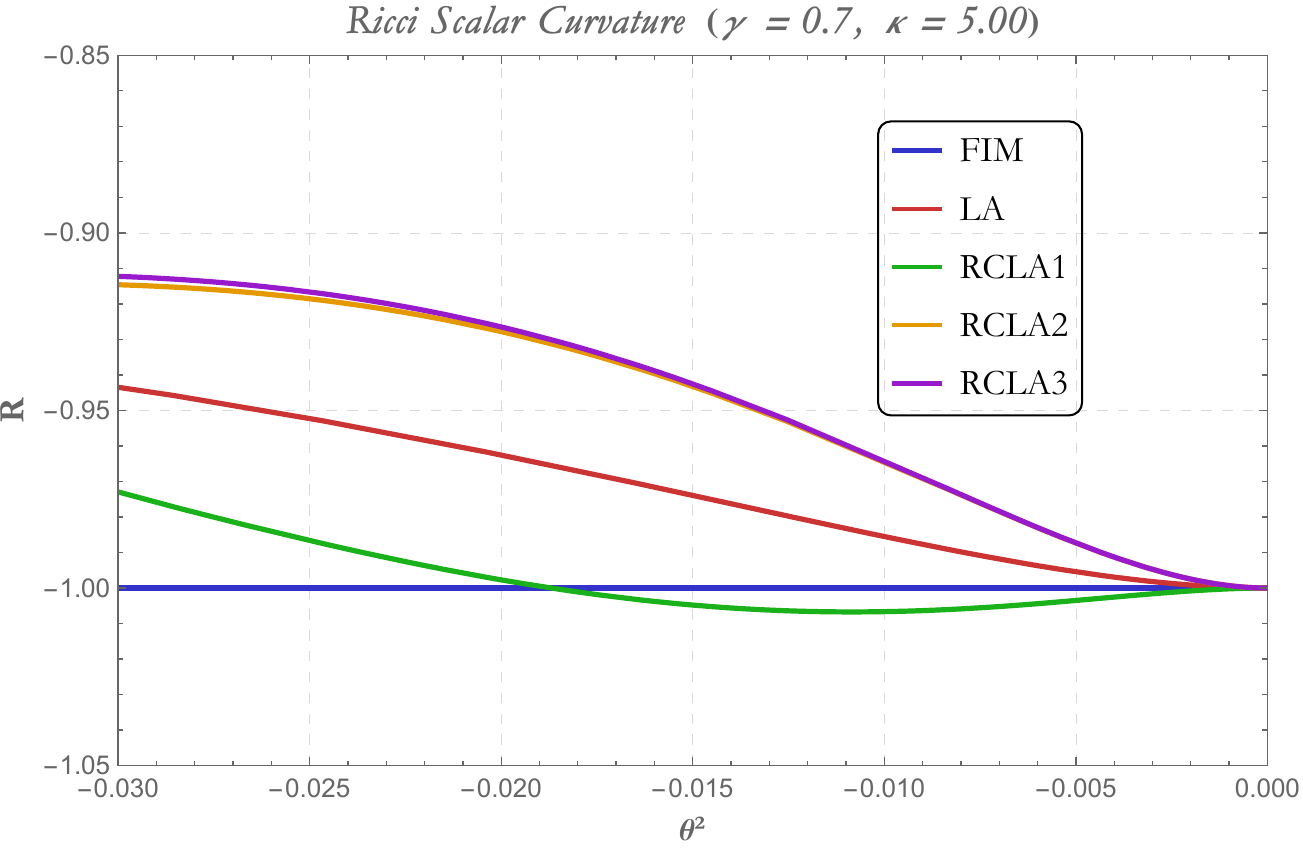}
		\caption{Variation of the Ricci scalar curvatures with  the coordinate $\theta^{2}$ for the five families of metrics considered. The geometry induced by all the metrics is still hyperbolic near the manifold boundary.}
\label{fig:SCALAR_CURAVTURE_THETA2}
	\end{minipage}
	\hspace{1.5 cm}
	\begin{minipage}{0.4\linewidth}
		\centering
\includegraphics[width=2.8in,height=2.5in]{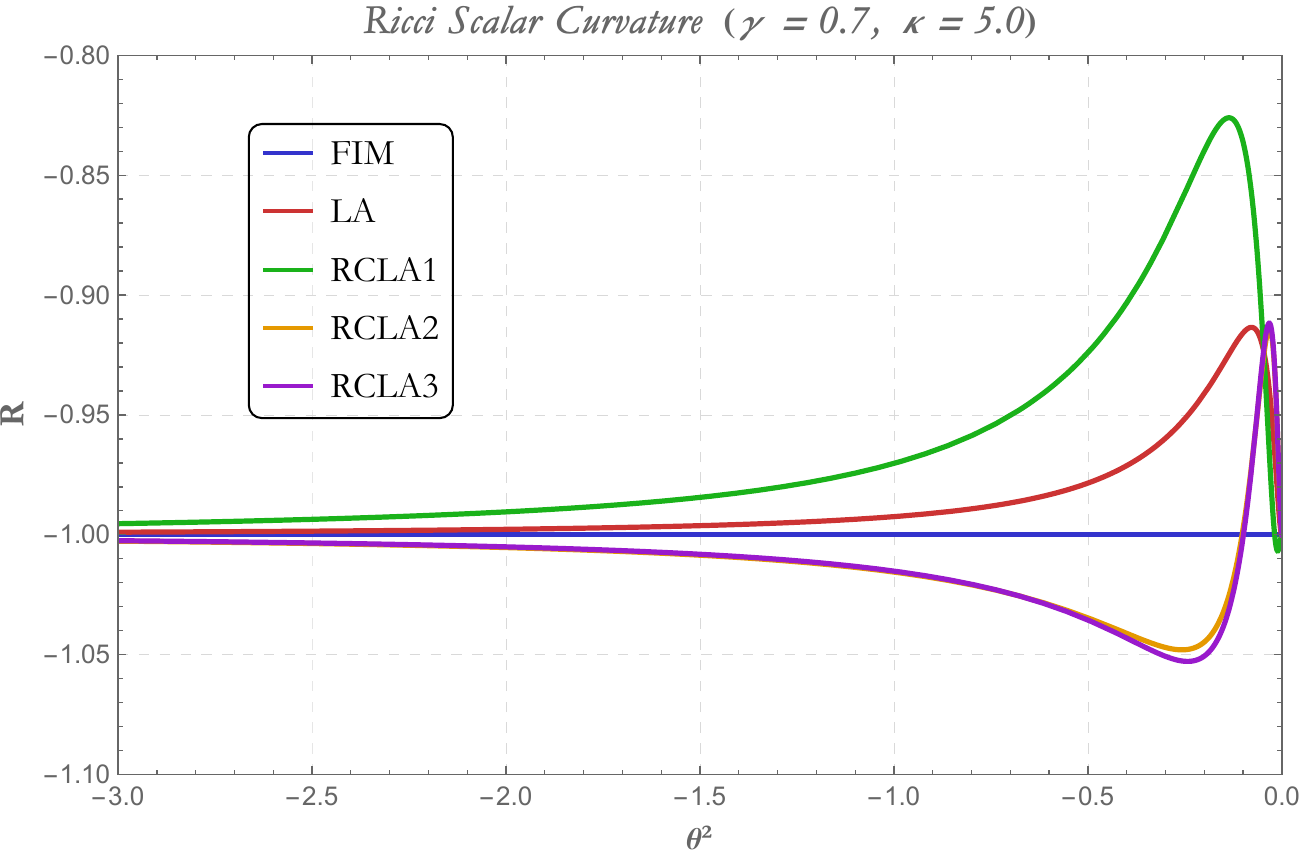}
		\caption{Variation of the Ricci scalar curvatures for larger values of  the coordinate $\theta^{2}$ for the five families of metrics considered. All of them converges to the FIM limit in this case.}
\label{fig:SCALAR_CURAVTURE_THETA2-LARGE}
	\end{minipage}
\end{figure}

Instead, we have shown a detailed comparison of the behaviour of each type of geometries with the coordinates $\theta^{2}$ and the control parameters $\gamma, \kappa$. In figures  \ref{fig:SCALAR_CURAVTURE_THETA2} and  \ref{fig:SCALAR_CURAVTURE_THETA2-LARGE}, we have shown how the Ricci scalar curvature $R$ varies with  the coordinate $\theta^{2}$ near the two boundaries. In particular, we want to point out that the geometry in all the cases is still hyperbolic near the manifold boundary $\theta^{2}=0$, as well as for $\theta^{2}\rightarrow -\infty$, where it approaches the FIM limit.

\begin{figure}[h!]
	\begin{minipage}{0.4\linewidth}
		\centering
\includegraphics[width=2.8in,height=2.5in]{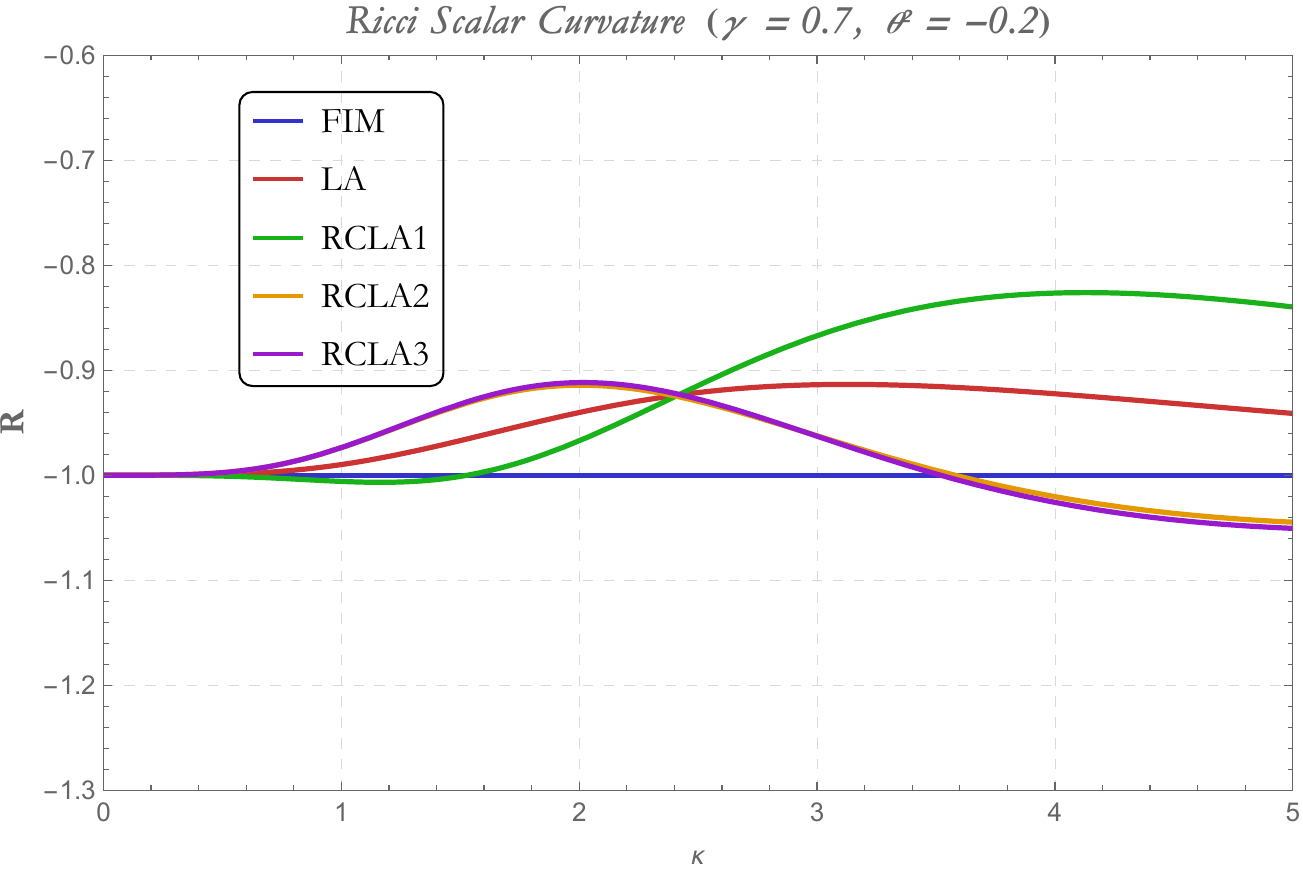}
		\caption{Variation of the Ricci scalar curvatures with the width of the Gaussian kernel $\kappa$, with $\gamma \leq 1$ for the five families of metrics considered.}
\label{fig:SCALAR_CURAVTURE_KAPPA}
	\end{minipage}
	\hspace{1.5 cm}
	\begin{minipage}{0.4\linewidth}
		\centering
\includegraphics[width=2.8in,height=2.5in]{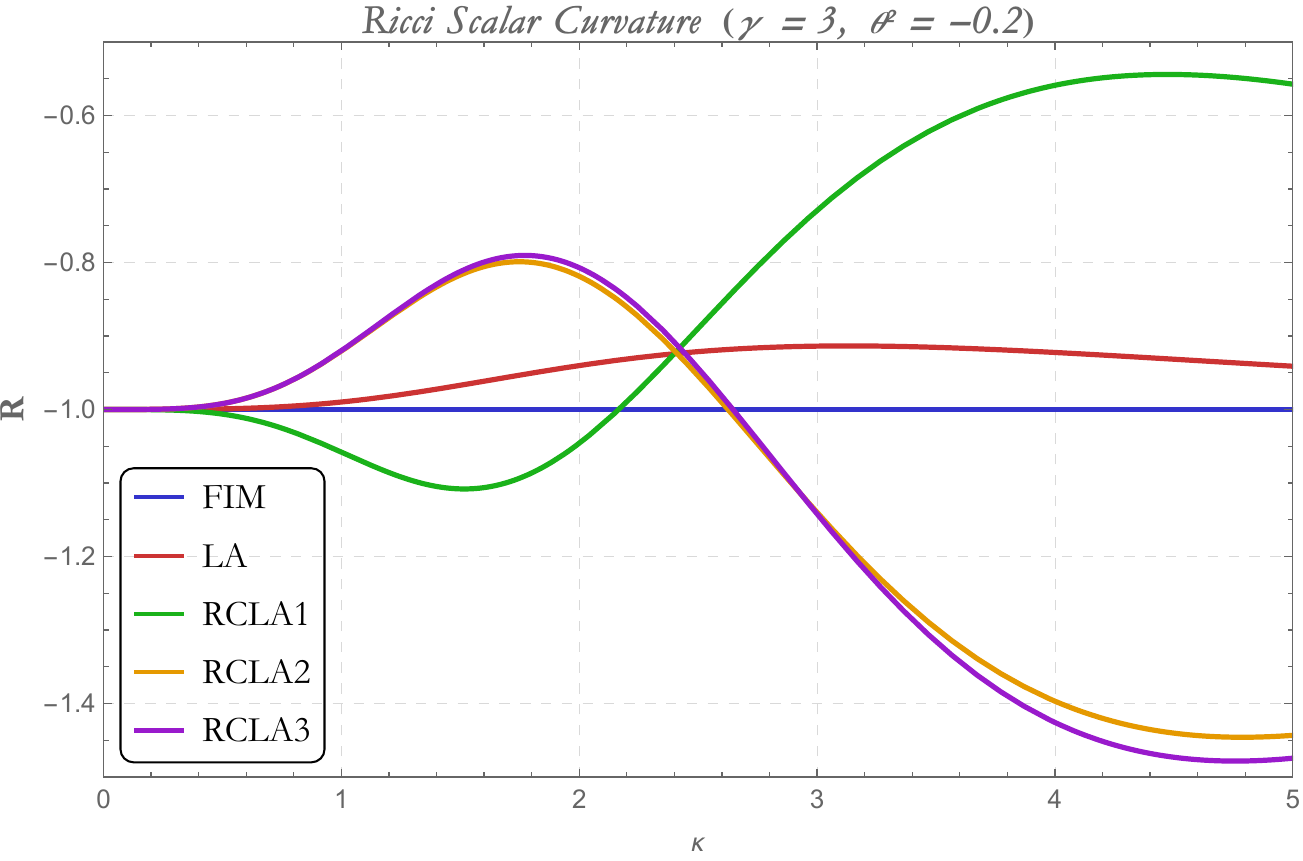}
		\caption{Variation of the Ricci scalar curvatures with the width of the Gaussian kernel $\kappa$, with $\gamma \geq 1$ for the five families of metrics considered.}
\label{fig:SCALAR_CURAVTURE_KAPPA_LARGE_KAPPA}
	\end{minipage}
\end{figure}

In figures \ref{fig:SCALAR_CURAVTURE_KAPPA} and \ref{fig:SCALAR_CURAVTURE_KAPPA_LARGE_KAPPA} we have plotted the curvature scalar variation with the width of the Gaussian kernel $\kappa$.  For values of $\gamma \leq 1$, where the bound on the effective learning rate is tighter (see sec. \ref{sec:CLAgeometry}) and for $\gamma >1$ as well.   

%%%%%%%%%%%
\bibliography{reference}%\bibliography{refs}
\end{document}